\documentclass[reprint,amsmath,amssymb,aps,prl,superscriptaddress]{revtex4-2}
\usepackage{CJK}
\usepackage[pdftex,colorlinks,linkcolor=purple,citecolor=red,urlcolor=blue]{hyperref}
\usepackage{mathrsfs}
\usepackage{mathtools}
\usepackage{rotating}
\usepackage{shuffle}
\usepackage{subcaption}
\usepackage{tikz}
\usepackage{tikz-feynman}
\usepackage{verbatim}

\allowdisplaybreaks

\let\ve=\varepsilon
\newcommand\imag{\text{i}}
\newcommand\mc{\omega}

\newcommand\mfd{\mathfrak{d}}
\newcommand\mfu{\mathfrak{u}}

\DeclareMathOperator\loceq{\circeq}

\makeatletter
\newcommand*{\balancecolsandclearpage}{%
  \close@column@grid
  \clearpage
  \twocolumngrid}
\makeatother

\newcommand{\rev}[1]{#1}
\newcommand{\del}[1]{}

\begin{document}
\begin{CJK*}{UTF8}{gbsn}

\title{Diagrammatic Derivation of Hidden Zeros and \\
Exact Factorisation of Pion Scattering Amplitudes}

\author{Yang Li (李阳)}\email{y.l.li@rug.nl}
\affiliation{Van Swinderen Institute for Particle Physics and Gravity, University of Groningen, 9747 Groningen, The Netherlands}

\author{Tianzhi Wang (王天志)}\email{tian.wang@rug.nl}
\affiliation{Van Swinderen Institute for Particle Physics and Gravity, University of Groningen, 9747 Groningen, The Netherlands}

\author{Tom\'a\v{s} Brauner}\email{tomas.brauner@uis.no}
\affiliation{Department of Mathematics and Physics, University of Stavanger, 4036 Stavanger, Norway}

\author{Diederik Roest}\email{d.roest@rug.nl}
\affiliation{Van Swinderen Institute for Particle Physics and Gravity, University of Groningen, 9747 Groningen, The Netherlands}

\begin{abstract}
Pion scattering amplitudes were recently found to vanish on specific kinematic loci, and to factorise close to these loci into a product of two lower-point amplitudes of an extended theory. We propose a diagrammatic representation of pion amplitudes that makes their vanishing on the loci manifest diagram by diagram. Moreover, we provide evidence that there is a closed-form expression for the amplitudes that generalises the near-zero factorisation in an exact manner, not only close to the loci but for all kinematic configurations. Our approach crucially relies on a novel formulation of the effective field theory of pions, in which tree-level scattering amplitudes are extracted from classical field equations for a set of covariantly conserved currents and emergent composite gauge fields.
\end{abstract}

\maketitle
\end{CJK*}


\emph{Introduction}---The theoretical study of pion physics combines a veritable past with an exciting present. In the massless limit, pions arise as Nambu-Goldstone (NG) modes during spontaneous symmetry breaking~\cite{Nambu:1960xd,Nambu:1961tp,Goldstone:1961eq,Goldstone:1962es} and are described by a nonlinear sigma model (NLSM). From the 1960s onwards, many interesting properties of their scattering amplitudes were unveiled, starting with the identification of the Adler zero in the soft limit~\cite{Adler:1964um}. In the subsequent decades, the NLSM appeared in many intriguing theoretical developments, including e.g.~the double-soft limit~\cite{Kampf:2013vha,Du:2015esa,Cheung:2016drk}, the double-copy prescription~\cite{Chen:2013fya,Cheung:2017ems,Cheung:2017pzi,Bern:2019prr}, and the Cachazo-He-Yuan formulation~\cite{Cachazo:2014xea,Low:2017mlh}. The latter allowed an identification of the leading soft behavior of the scattering amplitudes in the NLSM in terms of an extended theory dubbed NLSM+$\Phi^3$~\cite{Cachazo:2016njl,Low:2018acv}.

A crucial ingredient of the NLSM is the adjoint nature of pions under the global symmetry group. This allows for the decomposition of (tree-level) pion scattering amplitudes into so-called partial amplitudes with a specific flavour ordering~\cite{Elvang2015}. The cyclic nature of the $n$-point ($n$-pt) partial amplitude can be highlighted by representing the $n$ particles in the scattering process with $n$ points on a circle.
The kinematics of the scattering is captured in a Lorentz-invariant manner by the Mandelstam variables, $s_{ij\dotsb\ell}\equiv(p_i+p_j+\dots+p_\ell)^2$, where $p_i^\mu$, $i=1,\dotsc,n$ are the particle momenta, chosen by convention to have equal orientation with respect to the scattering zone.

A surprising feature of pion scattering was discovered in~\cite{Arkani-Hamed:2023swr} using the ``surfaceology'' formalism of~\cite{Arkani-Hamed:2023lbd,Arkani-Hamed:2023mvg}. (See also~\cite{Cao:2024gln,Bartsch:2024amu,Li:2024qfp,Cao:2024qpp} for related work.) Pick~two particles $1,k$ with $k\in\{3,\dotsc,n-1\}$. This splits the remaining $n-2$ pions into ``upper'' and ``lower'' subsets:\\[-3ex]
\begin{equation}
    \parbox{0.15\textwidth}{\includegraphics[width=0.15\textwidth]{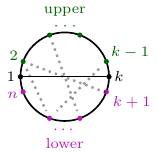}}\\[-1ex]
    \label{fig:zeroes}
\end{equation}
Then, define a particular kinematic locus as the subset of the kinematic space where all Mandelstam variables pairing one upper and one lower particle vanish,
 \begin{align}
     s_{ij} = 0 \quad \forall \,\, i\in\{2,\ldots,k-1\} \,, \,\, j\in\{k+1,\ldots,n\} \,.
     \label{eq: zero locus}
 \end{align}
At this locus, the partial amplitude for the ordering implied by~\eqref{fig:zeroes} vanishes. This property has become known as a \emph{hidden zero}. In the special case of $k=3$, it recovers the standard Adler zero. The latter itself can be understood as a consequence of the spontaneously broken symmetry of the NLSM. However, a similar symmetry argument for the more general hidden zeros is currently not available.

The first main result of this Letter is a novel framework for pion scattering amplitudes that makes the hidden zeros manifest. Namely, we develop a diagrammatic representation of tree-level amplitudes in which they vanish on the locus~\eqref{eq: zero locus} diagram by diagram. The basic ingredients of our approach are supplied by the Maurer-Cartan (MC) form descending directly from the spontaneously broken symmetry. These enter the classical equations of motion (EoM) of the NLSM, from which the amplitudes follow via Berends-Giele recursion~\cite{Berends:1987me}. 

As stressed in~\cite{Arkani-Hamed:2023swr}, the study of zeros provides a complementary view of scattering amplitudes, which are traditionally thought of in terms of poles instead; the latter define factorisation channels where the amplitude decomposes into a product of lower-point amplitudes. Remarkably, the kinematic loci~\eqref{eq: zero locus} also give rise to factorisation. Namely, in the near-zero limit, retaining only terms linear in the locus variables, the amplitude splits into a product of two lower-point amplitudes. For $k=3$, this builds on the identification of the leading contribution to the amplitude in the soft limit as the $(n-1)$-pt amplitude of the NLSM extended with a $\Phi^3$ coupling~\cite{Cachazo:2016njl}.

Our second main result is that, for any $k$, the factorisation defined by the kinematic loci~\eqref{eq: zero locus} can be made exact, to all orders in the locus variables, provided one subtracts a specific set of exchange diagram contributions. Our diagrammatic approach helps to make the exact factorisation property transparent~\footnote{An alternative diagrammatic approach to the hidden zeros and factorisation on the kinematic locus~\eqref{eq: zero locus} was put forward recently in~\cite{Zhou:2024ddy}.}. 



\emph{Covariant Formulation of NLSM}---We start with an effective theory of NG bosons on an arbitrary symmetric coset space $G/H$~\cite{Brauner2024}. The NG fields $\pi^a$, one for each broken symmetry generator $Q_a$, are encoded in a coset representative $U(\pi)\in G$. This leads to the MC one-form,
 \begin{equation}
   \mc_\mu\equiv-\imag U^{-1}\partial_\mu U\equiv  \mc^a_\mu Q_a + A^\alpha_\mu Q_\alpha \,,
   \label{mcform}
 \end{equation}
where $Q_\alpha$ is the set of (unbroken) generators of $H$. To determine the precise dependence of the MC form on the NG fields requires a specific choice of $U(\pi)$. One can however always ensure that upon power expansion in $\pi^a$,
\begin{equation}
    \mc^a_\mu=\partial_\mu\pi^a+\mathcal O(\pi^2\partial_\mu\pi)\,,\qquad
    A^\alpha_\mu=\mathcal O(\pi\partial_\mu\pi)\,. \label{lowest}
\end{equation}
Under symmetry transformations from $G$, $\mc^a_\mu$ behaves as a covariant vector field, whereas $A^\alpha_\mu$ is a composite $H$-valued connection. The latter can be used to construct the covariant derivative $D_\mu\mc^a_\nu$, and the field strength $F^\alpha_{\mu\nu}$, in the usual manner. These satisfy the constraints 
\begin{equation}
D_\mu\mc^a_\nu-D_\nu\mc^a_\mu=0\,, \qquad F^\alpha_{\mu\nu} = f^\alpha_{ab}\mc^a_\mu\mc^b_\nu \,,
\label{covdercomm}
\end{equation}
stemming from the MC structure equation, $\partial_\mu\mc_\nu-\partial_\nu\mc_\mu+\imag[\mc_\mu,\mc_\nu]=0$, where $f^\alpha_{ab}$ are the structure constants of~$G$.

At the leading order of the derivative expansion of the effective theory, the EoM for the NG fields $\pi^a$ amounts to the covariant conservation law \rev{$D_\mu\mc^{a\mu}=0$. Combining this with~\eqref{covdercomm}, we arrive at second-order equations} for the covariant ``current'' $\mc^a_\mu$ and the composite connection $A^\alpha_\mu$,
\begin{equation}
\begin{aligned}
  D^2\mc^a_\mu+f^a_{\alpha b}f^\alpha_{cd}\mc^b_\nu\mc^{c\nu}\mc^d_\mu&=0\,,\\
  D_\mu F^{\alpha\mu\nu} + f^\alpha_{ab}(D^\nu\mc^a_\mu)\mc^{b\mu}&=0 \,.\label{eomquartic}
\end{aligned}
\end{equation}
These equations constitute the backbone of our approach to the NLSM. Their key features are that they are explicitly agnostic of the choice of parametrisation of the coset space and, unlike the standard Lagrangian formulation of the effective theory of NG bosons, only include a finite number of interactions. While \eqref{eomquartic} holds for any symmetric coset space, we will from now on specialise to the chiral coset spaces $G/H\simeq\text{SU}(N)_\text{L}\times\text{SU}(N)_\text{R}/\text{SU}(N)_\text{V}$ as appropriate for the NLSM describing pion physics.



\emph{Tree Diagrams and Effective Vertices}---The classical equations~\eqref{eomquartic} can be used to generate tree-level pion amplitudes in a Berends-Giele fashion. Namely, solving the equations iteratively in presence of sources $J^a$ for asymptotic one-pion states gives the one-point function $\langle\mc^a_\mu\rangle_J$. This can be converted to on-shell amplitudes following the Lehmann-Symanzik-Zimmermann (LSZ) reduction.

The rules of the game are as follows~\cite{Monteiro:2011pc,Cheung:2021zvb}. Each term contributing to $\langle\mc^a_\mu\rangle_J$ can be represented by a rooted tree graph, with one external leg (\emph{root}) corresponding to $\mc^a_\mu$ itself. All the other $n-1$ legs (\emph{leaves}) carry the external source. The branching of the graphs is controlled by the nonlinear terms in~\eqref{eomquartic}, whose Feynman rules are summarised in Appendix A. Upon LSZ reduction, each external leg carries an on-shell momentum $p^\mu_i$. To connect correctly the composite field $\mc^a_\mu$ to asymptotic one-particle states, each external leg has to be dressed with a polarisation vector $\ve^\mu_i(p_i)$. The polarisations of the root and leaf legs are, respectively,
\begin{equation}
\ve^\mu_\text{root}(p)=-\imag \frac{ q^\mu}{q\cdot p}\,,\qquad
\ve^\mu_\text{leaf}(p)=\imag p^\mu\,.
\label{polarization}
\end{equation}
Here $q^\mu$ is an arbitrarily chosen auxiliary vector. This enters individual diagrams but will drop out of the final result for the on-shell amplitude. Finally, to agree with conventions used in the literature (see e.g.~\cite{Kampf:2013vha}), we add an overall minus sign for amplitudes of all multiplicities.

Importantly, the composite gauge field $A^\alpha_\mu$ does not couple to asymptotic one-pion states but only enters the amplitudes through internal propagators. All external legs of relevant graphs are of the $\mc^a_\mu$ type. This allows one to package certain sets of (sub)diagrams into effective vertices of even multiplicity. Thus, at 4-pt we~define 
\begin{equation}
\raisebox{-7 mm}{\scalebox{0.7}{
\begin{tikzpicture}
  \begin{feynman}[small]
    \vertex (a) {\(1\)};
    \node (v1) [circle, draw, right= 1.2 cm of a]{$V_{4}$};
    \vertex [right = 1.2cm of v1] (c) {\(3\)};
    \vertex [below right =  1.4cm of v1] (d) {\(4\)};
    \vertex [above right = 1.4cm of v1] (b) {\(2\)};
    \diagram* {
      (a) -- [solid,line width=0.5mm] (v1) --[solid,line width=0.5mm] (b),
      (v1) -- [solid,line width=0.5mm] (d),
      (v1) -- [solid,line width=0.5mm] (c)
    };
  \end{feynman}
\end{tikzpicture}}}
\equiv
\raisebox{-6 mm}{\scalebox{0.7}{
\begin{tikzpicture}
  \begin{feynman}[small]
    \vertex (a) {\(1\)};
    \vertex (v1) [right= 0.8 cm of a];
    \vertex [right = 0.8cm of v1] (c) {\(3\)};
    \vertex [below right =  0.8cm of v1] (d) {\(4\)};
    \vertex [above right = 0.8cm of v1] (b) {\(2\)};
    \diagram* {
      (a) -- [solid,line width=0.5mm] (v1) --[solid,line width=0.5mm] (b),
      (v1) -- [solid,line width=0.5mm] (d),
      (v1) -- [solid,line width=0.5mm] (c)
    };
  \end{feynman}
\end{tikzpicture}}}
+
\raisebox{-6 mm}{\scalebox{0.7}{
\begin{tikzpicture}
  \begin{feynman}[small]
    \vertex (a) {\(1\)};
    \vertex (v1) [right= 0.8 cm of a];
    \vertex (v2) [below right= 0.8 cm of v1];
    \vertex [above right = 0.6 cm of v2] (c) {\(3\)};
    \vertex [below right =  0.6 cm of v2] (d) {\(4\)};
    \vertex [above right = 0.8 cm of v1] (b) {\(2\)};
    \diagram* {
      (a) -- [solid,line width=0.5mm] (v1)-- [boson,line width=0.5mm] (v2) --[solid,line width=0.5mm] (d),
      (v1) -- [solid,line width=0.5mm] (b),
      (v2) -- [solid,line width=0.5mm] (c)
    };
  \end{feynman}
\end{tikzpicture}}}
+\raisebox{-6 mm}{\scalebox{0.7}{
\begin{tikzpicture}
  \begin{feynman}[small]
    \vertex (a) {\(1\)};
    \vertex (v1) [right= 0.8 cm of a];
    \vertex (v2) [above right= 0.8 cm of v1];
    \vertex [above right = 0.6 cm of v2] (c) {\(2\)};
    \vertex [below right =  0.6 cm of v2] (d) {\(3\)};
    \vertex [below right = 0.8 cm of v1] (b) {\(4\)};
    \diagram* {
      (a) -- [solid,line width=0.5mm] (v1)-- [boson,line width=0.5mm] (v2) --[solid,line width=0.5mm] (d),
      (v1) -- [solid,line width=0.5mm] (b),
      (v2) -- [solid,line width=0.5mm] (c)
    };
  \end{feynman}
\end{tikzpicture}}}\notag
\end{equation}
with solid lines representing the covariant current $\mc^a_\mu$ and wavy lines the composite connection $A^\alpha_\mu$. Converting from structure constants of the chiral symmetry group to the trace basis with the associated flavour ordering (see e.g.~\cite{Cheung:2017pzi}), we get an effective rule for the 4-pt vertex,
\begin{equation}
     \begin{aligned}
 V_4(1,2,3,4) = & - (1 + 2 p_2 \cdot (p_3-p_4) / s_{34} ) \delta_{12} \delta_{34}   \\
  & + \delta_{13} \delta_{24} + \text{mirror} \,.\label{V4}
\end{aligned}
\end{equation}
We used the shorthand notation $\delta_{ij}\equiv \eta_{\mu_i\mu_j}$ where $\mu_i$ are Lorentz indices attached to the external legs. ``Mirror'' indicates an expression obtained by inversion of the circle~\eqref{fig:zeroes} with respect to $i=1$, that is $i\to(1-i\mod n)+1$. 

The 4-pt partial amplitude, $A_4$, now follows from multiplying \eqref{V4} with the polarisation vectors~\eqref{polarization} for external particles and taking the on-shell limit. This results in 
 \begin{equation}
     A_4 = s_{24} \frac{q \cdot (p_2 + p_3 + p_4)}{q \cdot p_1}=-s_{24}\,.
     \label{A4}
 \end{equation}

Moving on to 6-pt, there will be diagrams that can be composed of two $V_4$-type vertices. The remaining 6-pt diagrams are precisely those that only contain $A^\alpha_\mu$-type propagators. We call such diagrams \emph{one-pion-irreducible} (1$\pi$I). Collecting all the 1$\pi$I 6-pt graphs gives a new effective contact 6-pt vertex,
\begin{align}
\notag
& V_6(1, \ldots, 6)=\\
\notag
 & - \frac{\delta_{12}\delta_{34}\delta_{56}}{s_{34}s_{56}} 
 \left(
    \begin{aligned}
     &-(p_3 - p_4)\cdot(p_5 - p_6)+\tfrac{2}{s_{3456}}\times\\
     &\quad
     \left(
     \begin{aligned}
     &+ p_2 \cdot p_3  (p_3 + 3 p_4) \cdot (p_5 - p_6) \\
     &- p_2 \cdot p_4  (3 p_3 + p_4) \cdot (p_5 - p_6) \\
     &- p_2 \cdot p_5  (p_3 - p_4) \cdot (p_5 + 3 p_6) \\
     &+ p_2 \cdot p_6  (p_3 - p_4) \cdot (3 p_5 + p_6)
     \end{aligned}
     \right)
    \end{aligned}
 \right)\\
\notag
 & +\frac{2\delta_{12}}{s_{3456}}
 \left(
 \begin{aligned}
     &+2\delta_{36}\delta_{45} p_2\cdot (p_4-p_5)/s_{45} \\
     &- \delta_{34}\delta_{56}p_2\cdot (p_3-p_4)/s_{34} \\
     &- \delta_{34}\delta_{56} p_2 \cdot (p_5-p_6)/s_{56}
 \end{aligned} 
 \right) \\
 & - \frac{\delta_{14}\delta_{23}\delta_{56}}{s_{23}s_{56}}(p_2 - p_3)\cdot(p_5 - p_6) + \text{mirror}
\label{V6}\,.
\end{align}
Together with exchange diagrams built from the 4-pt effective vertex~\eqref{V4} this recovers, after attaching the external polarisation vectors and going on-shell, the correct 6-pt partial amplitude of the NLSM.

The same logic carries over to higher multiplicities. For any even $n$, there is a class of diagrams that can be built by combining the lower-point effective vertices $V_4,V_6,\dotsc,V_{n-2}$. These are complemented by the 1$\pi$I $n$-pt diagrams, which define a new effective contact vertex, $V_n$. See the Supplemental Material~\cite{SM} for \rev{a compilation of the effective vertices up to and including $V_8$}.



\emph{Hidden Zeros from Diagrams}---\del{The approach to the NLSM based on~\eqref{eomquartic} is suited for illuminating specific features of pion scattering. To start with, note that}Due to the definition of the root polarisation vector in~\eqref{polarization}, the numerator of any amplitude is a linear combination of $q\cdot p_i$ where $i=2,\dotsc,n$. The $q$-independence of the final result for the on-shell amplitude requires that all the contributions fold into $q\cdot(p_2+\dotsb+p_n)$ so as to cancel the $q\cdot p_1$ factor in the denominator; see~\eqref{A4} for an explicit illustration. This means that the entire amplitude can be reconstructed by computing solely contributions proportional to a fixed $q\cdot p_k$ factor. We call the collection of such contributions a \emph{splitting}, $(1\to k)$, of the amplitude.

\del{The freedom to choose a splitting can be leveraged to simplify the calculation of the amplitude and highlight its properties. Indeed, }We will now show that choosing a fixed splitting $(1\to k)$ manifests the vanishing of the amplitude on the kinematic locus~\eqref{eq: zero locus} defined by the same $k$. The story will be clearest for odd $k$; we will address this case first and then comment on even $k$. In both cases, we will focus on simple illustrative examples, relegating details of the general argument to appendices.

At 6-pt, there is only one type of odd splitting, $(1 \to 3)$, up to cyclicity implied by~\eqref{fig:zeroes}. This splitting does not get any contribution from the contact vertex~\eqref{V6}, and is determined entirely by diagrams of the $V_4\otimes V_4$ type,
\begin{equation}
\begin{small} 
\raisebox{-9 mm}{\scalebox{0.7}{
\begin{tikzpicture}
  \begin{feynman}
    \vertex (a) {\(1\)};
    \node (v1) [circle, draw, right= 1.2 cm of a]{$V_{4}$};
    \node (v3) [circle, draw, right= 1.2 cm of v1]{$V_{4}$};
    \vertex [right= 1.2cm of v3] (e) {\(3\)};
    \vertex [below left = 1.2cm of v3] (c) {\(5\)};
    \vertex [below right =  1.2cm of v3] (d) {\(4\)};
    \vertex [above = 1.2cm of v1] (b) {\(2\)};
    \vertex [below = 1.2cm of v1] (f) {\(6\)};
    \diagram* {
      (a) -- [double,line width=0.3mm] (v1) --[double,line width=0.3mm] (v3) -- [double,line width=0.3mm] (e),
      (v1) -- [dashed,line width=0.5mm] (b),
      (v1) -- [dashed,line width=0.5mm] (f),
      (v3) -- [solid,line width=0.5mm] (d),
      (v3) -- [solid,line width=0.5mm] (c)
    };
  \end{feynman}
\end{tikzpicture}}}
+
\raisebox{-9 mm}{\scalebox{0.7}{
\begin{tikzpicture}
  \begin{feynman}
    \vertex (a) {\(1\)};
    \node (v1) [circle, draw, right= 1.2 cm of a]{$V_{4}$};
    \node (v3) [circle, draw, right= 1.2 cm of v1]{$V_{4}$};
    \vertex [below left = 1.2 cm of v1] (b) {\(6\)};
    \vertex [below right = 1.2 cm of v1] (c) {\(5\)};
    \vertex [below = 1.2 cm of v3] (f) {\(4\)};
    \vertex [above = 1.2 cm of v3] (d) {\(2\)};
    \vertex [right = 1.2 cm of v3] (e) {\(3\)};
    \diagram* {
      (a) -- [double,line width=0.3mm] (v1) --[double,line width=0.3mm] (v3) -- [double,line width=0.3mm] (e),
      (v3) -- [dashed,line width=0.5mm] (d),
      (v3) -- [dashed,line width=0.5mm] (f),
      (v1) -- [solid,line width=0.5mm] (b),
      (v1) -- [solid,line width=0.5mm] (c)
    };
  \end{feynman}
\end{tikzpicture}}}
+
\raisebox{-9 mm}{\scalebox{0.7}{ 
  \begin{tikzpicture}
  \begin{feynman}[small]
    \vertex (a) {\(6\)};
    \node (v2) [circle, draw, right= 1.2 cm of a]{$V_{4}$};
    \vertex [right= 1.2cm of v2] (e) {\(4\)} ;
    \node (v4) [circle, draw, above= 1.2 cm of v2]{$V_{4}$};
    \vertex [right = 1.2 cm of v4] (d) {\(3\)};
    \vertex [left = 1.2 cm of v4] (b) {\(1\)};
    \vertex [above =1.2 cm of v4] (c) {\(2\)};
    \vertex [below = 1.2 cm of v2] (f) {\(5\)};
    \diagram* {
      (a) -- [dashed,line width=0.5mm] (v2) -- [dashed,line width=0.5mm] (e),
      (v2) -- [dashed,line width=0.5mm] (v4) -- [dashed,line width=0.5mm] (c),
      (v4) -- [double,line width=0.3mm] (b),
      (v4) -- [double,line width=0.3mm] (d),
      (v2) -- [dashed,line width=0.5mm] (f)
    };
  \end{feynman}
\end{tikzpicture}}}
\label{eq: 6-pt ac zero}
\end{small}
\end{equation}
where the doubled line indicates the splitting. The contributions of the three different diagrams are straightforward to evaluate using~\eqref{V4}, which leads to 
\begin{equation}
\begin{small}
\begin{aligned}
    A_6^{(1\to3)}={}&\frac{ s_{26} (s_{34} + s_{45} -s_{35} )}{2s_{126}} + \frac{ s_{24} (s_{56} + s_{16} -s_{15} )}{2s_{156}}\\ 
   &+ \frac{(s_{24} + s_{26}) (s_{45} + s_{56} - s_{46} ) - 2 s_{25} s_{46}}{2s_{123}} \,,
   \end{aligned}
   \end{small}
\label{eq: 6-pt ac_1}
\end{equation}
where the superscript $(1\to k)$ on the amplitude indicates the chosen splitting. 

Remarkably, each term in~\eqref{eq: 6-pt ac_1} vanishes separately on the kinematic locus~\eqref{eq: zero locus} defined by the $(1\to3)$ splitting. The reason for this is that each diagram in~\eqref{eq: 6-pt ac zero} includes a 4-pt vertex that is traversed ``diagonally'' by the splitting. The 4-pt vertex \eqref{V4} then implies that, upon dressing with polarisation vectors, the diagram is necessarily proportional to some of the Mandelstam variables that define the $(1\to3)$ locus, as indicated by the dashed lines in~\eqref{eq: 6-pt ac zero}. Further illustrative examples are offered by the 8-pt amplitude; we provide a detailed check of the vanishing of the 8-pt amplitude on all the loci~\eqref{eq: zero locus} at the effective diagram level in the Supplemental Material~\cite{SM}.

The argument is readily generalised to all odd splittings for amplitudes of any multiplicity. To reach the $k$-leaf from the root, the $(1\to k)$ splitting must collect an odd number of $\delta_{1i}$ factors with odd $i$ as it traverses the individual effective vertices. However, the only such factor appears in the $\delta_{13}$ term of~\eqref{V4}. (A detailed proof of the absence of any other $\delta_{1i}$ factors with odd $i$ is provided in Appendix B.) Hence, an odd splitting must traverse diagonally an odd number of 4-pt vertices. This implies that any diagram contributing to the $(1\to k)$ splitting will be at least linear in the Mandelstam variables that define the $(1\to k)$ locus. Thus, for odd splittings, the hidden zeros of~\cite{Arkani-Hamed:2023swr} are manifest diagram by diagram.

The situation is more subtle for even splittings. We will illustrate this on the sole nontrivial even splitting, $(1\to 4)$, of the 6-pt amplitude, which takes the form
\begin{equation}
\begin{small}
\begin{aligned}
&\raisebox{-8 mm}{\scalebox{0.8}{
\begin{tikzpicture}
  \begin{feynman}[small]
    \vertex (a) {\(1\)};
    \node (v1) [circle, draw, right= 1.2 cm of a]{$V_{4}$};
    \node (v2) [circle, draw, right= 1.2 cm of v1]{$V_{4}$};
    \vertex [above = 1.0cm of v1] (b) {\(2\)};
    \vertex [right=of v2] (d) {\(4\)};
    \vertex [above = 1.0cm of v2] (c) {\(3\)};
    \vertex [below = 1.0cm of v2] (e) {\(5\)};
    \vertex [below = 1.0cm of v1] (f) {\(6\)};
    \diagram* {
      (a) -- [double,line width=0.3mm] (v1) -- [double,line width=0.3mm] (v2) -- [double,line width=0.3mm] (d),
      (v2) -- [dashed,line width=0.5mm] (e),
      (v2) -- [dashed,line width=0.5mm] (c),
      (v1) -- [dashed,line width=0.5mm] (b),
      (v1) -- [dashed,line width=0.5mm] (f)
    };
  \end{feynman}
\end{tikzpicture}}}
+
\raisebox{-8 mm}{\scalebox{0.8}{
\begin{tikzpicture}
  \begin{feynman}[small]
    \vertex (a) {\(1\)};
    \node (v1) [circle, draw, right= 1.2 cm of a]{$V_{4}$};
    \node (v3) [circle, draw, right= 1.2 cm of v1]{$V_{4}$};
    \vertex [above left = 1.2cm of v3] (b) {\(2\)};
    \vertex [right = 1.2cm of v3] (d) {\(4\)};
    \vertex [above right = 1.2cm of v3] (c) {\(3\)};
    \vertex [below right = 1.2cm of v1] (e) {\(5\)};
    \vertex [below left =1.2cm of v1] (f) {\(6\)};
    \diagram* {
      (a) -- [double,line width=0.3mm] (v1) --[double,line width=0.3mm] (v3) -- [double,line width=0.3mm] (d),
      (v1) -- [solid,line width=0.5mm] (e),
      (v1) -- [solid,line width=0.5mm] (f),
      (v3) -- [solid,line width=0.5mm] (b),
      (v3) -- [solid,line width=0.5mm] (c)
    };
  \end{feynman}
\end{tikzpicture}}}\\
&
+
\raisebox{-8 mm}{\scalebox{0.8}{
\begin{tikzpicture}
  \begin{feynman}[small]
    \vertex (a) {\(1\)};
   \node (v1) [circle, draw, right= 1.2 cm of a]{$V_{6}$};
    \vertex [above left = 1.2 of v1] (b) {\(2\)};
    \vertex [above right = 1.2 of v1] (c) {\(3\)};
    \vertex [below left = 1.2 of v1] (f) {\(6\)};
    \vertex [below right = 1.2 of v1] (e) {\(5\)};
    \vertex [right = 1.2 of v1] (d) {\(4\)};
    \diagram* {
      (a) -- [double,line width=0.3mm] (v1) -- [double,line width=0.3mm] (d),
      (v1) -- [dashed,line width=0.5mm] (e),
      (v1) -- [dashed,line width=0.5mm] (f),
      (v1) -- [dashed,line width=0.5mm] (b),
      (v1) -- [dashed,line width=0.5mm] (c)
    };
  \end{feynman}
\end{tikzpicture}}}
\quad + 
\raisebox{-8 mm}{\scalebox{0.8}{
\begin{tikzpicture}
  \begin{feynman}[small]
    \vertex (a) {\(1\)};
     \node (v1) [circle, draw, right= 1.2 cm of a]{$V_{4}$};
      \node (v3) [circle, draw, right= 1.2 cm of v1]{$V_{4}$};
    \vertex [above left = 1.2 cm of v1] (b) {\(2\)};
    \vertex [above right = 1.2 cm of v1] (c) {\(3\)};
    \vertex [below left = 1.2 cm of v3] (f) {\(6\)};
    \vertex [below right = 1.2 cm of v3] (e) {\(5\)};
    \vertex [right = 1.2 cm of v3] (d) {\(4\)};
    \diagram* {
      (a) -- [double,line width=0.3mm] (v1) -- [double,line width=0.3mm] (v3) -- [double,line width=0.3mm] (d),
      (v3) -- [solid,line width=0.5mm] (e),
      (v3) -- [solid,line width=0.5mm] (f),
      (v1) -- [solid,line width=0.5mm] (b),
      (v1) -- [solid,line width=0.5mm] (c)
    };
  \end{feynman}
\end{tikzpicture}}}\\ 
=&-\frac{ s_{26} s_{35} }{s_{126}} - \frac{ (s_{56} + s_{16} -s_{15} ) (s_{23} + s_{34} -s_{24} ) }{4s_{156}} - \\ 
&\frac{s_{25} + s_{36} -s_{26}-s_{35}}{4} - \frac{ (s_{12} + s_{23} -s_{13} ) (s_{45} + s_{56} -s_{46} ) }{4s_{123}}\,.
\end{aligned}\label{eq: three (4,4) diagrams}
\end{small}
\end{equation}
It is easy to check that this equals \eqref{eq: 6-pt ac_1}. The first and third contributions vanish on the $(1\to 4)$ locus. On the other hand, the second and fourth terms only vanish once combined together. To see why, note first that $0=s_4=s_{12356}\loceq s_{123}+s_{156}$, where $\loceq$ denotes equality on the kinematic locus. Thus, $s_{123}\loceq-s_{156}$, and the second and fourth terms in~\eqref{eq: three (4,4) diagrams} cancel if the numerators of the fractions equal each other. This follows from $s_{12}-s_{13}=s_{3456}-s_{2456}\loceq s_{34}-s_{24}$, and analogously $s_{45}-s_{46}=s_{1236}-s_{1235}\loceq s_{16}-s_{15}$. The pattern repeats for higher-multiplicity amplitudes and $(1\to k)$ splittings with even $k$. Some diagrams vanish on the $(1\to k)$ locus manifestly. Others do not vanish separately, but instead require cancellation within groups of diagrams of similar topologies; see Appendix C for further details.



\emph{Exact Factorisation}---\del{We have shown how to manifest the hidden zeros of NLSM amplitudes by a suitable choice of splitting. However, our approach provides intriguing evidence for a much stronger result. }\rev{Not only do NLSM amplitudes vanish on the kinematic locus~\eqref{eq: zero locus}, but they} can be expressed as a sequence of terms of increasing order in the locus variables. The constant part vanishes, and the linear part factorises into a product of two subamplitudes. The higher-order parts likewise factorise into well-defined building blocks. We start with examples, and then formulate a general conjecture.

We first introduce an ``extended amplitude,'' which plays a key role in the factorisation for odd splittings. This is an object with an odd number of external legs, of which all but three carry pions $\pi^a$. The remaining three legs carry other scalars $\Phi^a$. The extended amplitude is then defined by the sum over all diagrams of the type
\begin{equation}
\parbox{3cm}{\raisebox{-8 mm}{\scalebox{0.7}{ 
\begin{tikzpicture}
  \begin{feynman}
    \vertex (1) {\({\Phi}\)};
    \node (v2) [circle, draw, right= 1.0 cm of 1]{$A_{m_1}$};
    \vertex (n1) [right =1 cm of v2];
    \vertex [below left = 0.5 cm and 0.7cm of v2] (n) {\begin{rotate}{-45}$  \cdots $\end{rotate}};
    \vertex [below right = 0.6 cm and 0.5 cm of v2] (2) {\begin{rotate}{45}$  \cdots $\end{rotate}};
    \vertex [above left = 0.4 cm and 0.6 cm of v2] (nn) {\begin{rotate}{45}$  \cdots $\end{rotate}};
    \vertex [above right = 0.7 cm and 0.3 cm of v2] (22) {\begin{rotate}{-45}$  \cdots $\end{rotate}};
    \vertex [below = 1 cm of v2] (x1) {\(\pi\)};
    \vertex [above = 1 cm of v2] (x11) {\(\pi\)};
    \node (vq) [circle, draw, below right= 0.7 cm of n1]{$A_{m_2}$};
    \vertex [below right = 1 cm of vq] (q2) {\({\Phi}\)};
    \vertex [below right = 0.2cm and 0.75 cm of vq] (x2) {\begin{rotate}{90}$  \cdots $\end{rotate}};
    \vertex [below left = 0.2cm and 0.6 cm of vq] (dotq1) {\begin{rotate}{90}$  \cdots $\end{rotate}};
    \vertex [above = 0.7 cm of vq] (dotq2) {\(\dots\)};
    \vertex [above right = 1 cm of vq] (q22) {\(\pi\)};
    \vertex [below left = 1 cm of vq] (q23) {\(\pi\)};
    \vertex [below = 0.7 cm of vq] (x22) {\(\dots\)};
    \node (vr) [circle, draw, above right= 0.7 cm of n1]{$A_{m_3}$};
    \vertex [above left = 1 cm of vr] (r2) {\(\pi\)};
    \vertex [below left = 0.2cm and 0.6 cm of vr] (dotr1) {\begin{rotate}{90}$  \cdots $\end{rotate}};
    \vertex [below = 0.7 cm of vr] (dotr2) {\(\dots\)};
    \vertex [above = 0.7 cm of vr] (x3) {\(\dots\)};
    \vertex [above right = 1 cm of vr] (r22) {\({\Phi}\)};
    \vertex [below right = 1 cm of vr] (r23) {\(\pi\)};
    \vertex [below right = 0.2 cm and 0.7 cm of vr] (x32) {\begin{rotate}{90}$  \cdots $\end{rotate}};
    +
    \diagram* {
      (1) -- [double,line width=0.3mm] (v2) -- [double,line width=0.3mm] (n1) -- [double,line width=0.3mm] (vq),
      (n1) -- [double,line width=0.3mm] (vr),
      (v2) -- [solid,line width=0.5mm] (x1),
      (v2) -- [solid,line width=0.5mm] (x11),
      (vq) -- [double,line width=0.3mm] (q2),
      (vq) -- [solid,line width=0.5mm] (q22),
      (vq) -- [solid,line width=0.5mm] (q23),
      (vr) -- [solid,line width=0.5mm] (r2),
      (vr) -- [solid,line width=0.5mm] (r23),
      (vr) -- [double,line width=0.3mm] (r22)
    };
  \end{feynman}
\end{tikzpicture}}}}
\label{fig:pqr}
\end{equation}
where the external legs are partitioned into three subsets of odd size, each of which contains one leaf leg $\Phi^a$. The mathematical representation of (partially off-shell) subamplitudes $A_{m_i}$ with a nonzero number of external pion legs is constructed using the same $V_n$-type effective vertices as before, with the choice of splitting $(1\to k)$ dictated by the  doubled lines, attaching polarisation vectors, and then extracting the coefficient of $(q\cdot p_k)/(q\cdot p_1)$. The root leg of each such a subamplitude is attached to the cubic vertex in the centre of~\eqref{fig:pqr} by an appropriate propagator. The central $\Phi^3$ vertex itself is constant and, by construction, does not satisfy momentum conservation. \rev{The latter feature accounts for the fact that the extended amplitude only depends on a subset of the momenta entering the full amplitude being factorised.}

The simplest example of an extended amplitude is the plain 3-pt vertex with no pion legs, $\mathcal{A}_3^{\text{ext}}(1^{\Phi},2^{\Phi},3^{\Phi})=1$, where the superscripts on the arguments indicate the type of external legs. Further simple examples include
 \begin{equation}
     \begin{split}
         &\mathcal{A}_5^\text{ext}(1^{\Phi},2^{\Phi},3^{\pi},4^\pi,5^{\Phi})=\frac{s_{24}}{s_{234}}+\frac{s_{35}}{s_{345}}-1\,,\\
         &\mathcal{A}_5^\text{ext}(1^{\Phi},2^{\pi},3^{\Phi},4^\pi,5^{\Phi})=\frac{s_{24}}{s_{234}}\,,\\
         &\mathcal{A}_5^\text{ext}(1^{\Phi},2^{\pi},3^{\pi},4^\Phi,5^{\Phi})=\frac{s_{24}}{s_{234}}+\frac{s_{13}}{s_{123}}-1\,.
     \end{split} \label{5pt mixed amplitude}
 \end{equation}
The above-defined extended amplitudes enter the decomposition of the full partial amplitudes of the NLSM in odd splittings $(1\to k)$. At 4-pt, this is rather trivial,
\begin{equation}
        A_{4}^{(1\to 3)}=-s_{24} \mathcal{A}_3^\text{ext}(1^{\Phi},2^{\Phi},3^{\Phi})\mathcal{A}_3^\text{ext}(3^{\Phi},4^{\Phi},1^{\Phi})\,. \label{A4-fact}
\end{equation}
The first nontrivial example appears at 6-pt. The partial amplitude~\eqref{eq: 6-pt ac_1} in the $(1\to3)$ splitting can be written as 
\begin{equation}
\begin{split}
    A_6^{(1\to 3)}=&-\mathcal{A}_3^\text{ext}(1^{\Phi},2^{\Phi},3^{\Phi})\\
    &\times\sum_{i=4,5,6} s_{2i}\mathcal{A}_5^\text{ext} (3^{\Phi},\dots,i^{\Phi},\dots,1^{\Phi})\,.\label{eq: 6-pt ac all}
\end{split}
\end{equation}
The individual contributions can be represented graphically as follows. Take the three diagrams in~\eqref{eq: 6-pt ac zero} and split them along the doubled line, interpreting the upper and lower parts as extended amplitudes. Taking into account all possibles choices for the external $\Phi$-legs, we thus get 
\begin{equation}
    \begin{small}
        \begin{aligned}
&\raisebox{-6 mm}{\scalebox{0.7}{
\begin{tikzpicture}
  \begin{feynman}[small]
    \vertex (a);
    \vertex [right= 0.7 cm of a] (v1);
    \vertex [right= 0.5 cm of v1] (v2) ;
    \vertex [right= 0.5 cm of v2] (v3) ;
    \vertex [above= 0.2 cm of v1] (vup1) ;
    \vertex [left= 0.67 cm of vup1] (aup) {\(1\)};
    \vertex [right= 1.7 cm of vup1] (cup) {\(3\)};
    \vertex [right= 0.7cm of v3] (e);
    \vertex [below = 0.7cm of v3] (c) {\(5\)};
    \vertex [below right =  0.7cm of v3] (d) {\(4\)};
    \vertex [above = 0.7cm of vup1] (b) {\(2\)};
    \vertex [below = 0.7cm of v1] (f) {\(6\)};
    \diagram* {
      (v1) -- [double,line width=0.3mm] (v2) --[double,line width=0.3mm] (v3) -- [double,line width=0.3mm] (e),
      (v1) -- [double,line width=0.3mm] (a),
      (vup1) -- [double,line width=0.3mm] (aup),
      (vup1) -- [dashed,line width=0.5mm] (b),
      (vup1) -- [double,line width=0.3mm] (cup),
      (v1) -- [dashed,line width=0.5mm] (f),
      (v3) -- [solid,line width=0.5mm] (d),
      (v3) -- [solid,line width=0.5mm] (c)
    };
  \end{feynman}
\end{tikzpicture}}}
+
\raisebox{-6 mm}{\scalebox{0.7}{
\begin{tikzpicture}
  \begin{feynman}[small]
    \vertex (a);
    \vertex [right= 0.7 cm of a] (v1);
    \vertex [right= 0.5cm of v1] (v2);
    \vertex [right=0.5 cm of v2] (v3) ;
    \vertex [above= 0.2 cm of v3] (vup) ;
    \vertex [right= 0.7 cm of vup] (cup) {\(3\)};
    \vertex [left= 1.67 cm of vup] (aup) {\(1\)} ;
    \vertex [below left = 0.7cm of v1] (b) {\(6\)};
    \vertex [below = 0.7cm of v1] (c) {\(5\)};
    \vertex [below = 0.7cm of v3] (f) {\(4\)};
    \vertex [above = 0.7cm of vup] (d) {\(2\)};
    \vertex [right = 0.7cm of v3] (e);
    \diagram* {
      (v3) -- [double,line width=0.3mm] (e),
      (v3) --[double,line width=0.3mm] (v2) -- [double,line width=0.3mm] (v1) -- [double,line width=0.3mm] (a),
      (vup) -- [dashed,line width=0.5mm] (d),
      (vup) -- [double,line width=0.3mm] (aup),
      (vup) -- [double,line width=0.3mm] (cup),
      (v3) -- [dashed,line width=0.5mm] (f),
      (v1) -- [solid,line width=0.5mm] (b),
      (v1) -- [solid,line width=0.5mm] (c)
    };
  \end{feynman}
\end{tikzpicture}}}
+
\raisebox{-8 mm}{\scalebox{0.7}{ 
  \begin{tikzpicture}
  \begin{feynman}[small]
    \vertex (a);
    \vertex [right= 0.4 cm of a] (v1);
    \vertex [right= 0.6 cm of v1] (v2);
    \vertex [above= 0.2 cm of v2] (vup);
    \vertex [right= 1 cm of vup] (cup) {\(3\)} ;
    \vertex [left= 1 cm of vup] (aup) {\(1\)};
    \vertex [right= 0.5cm of v2] (v3) ;
    \vertex [right= 0.5cm of v3] (e) ;
    \vertex [below = 0.5cm of v2] (v4);
    \vertex [right = 0.5cm of v4] (d) {\(4\)};
    \vertex [left = 0.5cm of v4] (b) {\(6\)};
    \vertex [below =0.5cm of v4] (c) {\(5\)};
    \vertex [above = 0.5cm of vup] (f) {\(2\)};
    \diagram* {
      (v2) --[double,line width=0.3mm] (v3) -- [double,line width=0.3mm] (e),
      (v2) -- [double,line width=0.3mm] (v1) -- [double,line width=0.3mm] (a),
      (v2) -- [dashed,line width=0.5mm] (v4) -- [dashed,line width=0.5mm] (c),
      (v4) -- [dashed,line width=0.5mm] (b),
      (v4) -- [dashed,line width=0.5mm] (d),
      (vup) -- [dashed,line width=0.5mm] (f),
      (vup) -- [double,line width=0.3mm] (aup),
      (vup) -- [double,line width=0.3mm] (cup)
    };
  \end{feynman}
\end{tikzpicture}}}
\\
=&
\raisebox{-5 mm}{\scalebox{0.7}{ 
  \begin{tikzpicture}
  \begin{feynman}[small]
    \vertex (a) {\(1\)};
    \vertex [right= 0.7cm of a] (v1);
    \vertex [right=0.4cm of v1] (e) {\(3\)} ;
    \vertex [above =0.4cm of v1] (f) {\(2\)};
    \diagram* {
      (a) -- [double,line width=0.3mm] (v1) -- [double,line width=0.3mm] (e),
      (v1) -- [double,line width=0.3mm] (f)
    };
  \end{feynman}
\end{tikzpicture}}}
\times
\left(
\begin{aligned}
&
\raisebox{-3 mm}{\scalebox{0.7}{ 
  \begin{tikzpicture}
  \begin{feynman}[small]
    \vertex (a) {\(1\)};
    \vertex [right=of a] (v1);
    \vertex [right=of v1] (v2);
    \vertex [right=of v2] (e) {\(3\)} ;
    \vertex [below right = 0.5cm of v2] (d) {\(4\)};
    \vertex [below = 0.5cm of v2] (c) {\(5\)};
    \vertex [below =0.5cm of v1] (b) {\(6\)};
    \diagram* {
      (a) -- [double,line width=0.3mm] (v1)-- [double,line width=0.3mm] (v2) -- [double,line width=0.3mm] (e),
      (v2) -- [solid,line width=0.5mm] (c),
      (v1) -- [double,line width=0.3mm] (b),
      (v2) -- [solid,line width=0.5mm] (d)
    };
  \end{feynman}
\end{tikzpicture}}}
+
\raisebox{-3 mm}{\scalebox{0.7}{ 
  \begin{tikzpicture}
  \begin{feynman}[small]
    \vertex (a) {\(1\)};
    \vertex [right=of a] (v1);
    \vertex [right=of v1] (v2);
    \vertex [right= 0.5 cm of v2] (e) {\(3\)} ;
    \vertex [below = 0.5cm of v1] (c) {\(5\)};
    \vertex [below left = 0.5cm of v1] (b) {\(6\)};
    \vertex [below =0.5cm of v2] (d) {\(4\)};
    \diagram* {
      (a) -- [double,line width=0.3mm] (v1)-- [double,line width=0.3mm] (v2) -- [double,line width=0.3mm] (e),
      (v1) -- [solid,line width=0.5mm] (c),
      (v1) -- [solid,line width=0.5mm] (b),
      (v2) -- [double,line width=0.3mm] (d)
    };
  \end{feynman}
\end{tikzpicture}}}
+\\
&\raisebox{-3 mm}{\scalebox{0.7}{ 
  \begin{tikzpicture}
  \begin{feynman}[small]
    \vertex (a) {\(1\)};
    \vertex [right=of a] (v1);
    \vertex [right= 0.5cm of v1] (e) {\(3\)} ;
    \vertex [below = 0.5cm of v1] (v4);
    \vertex [right = 0.3cm of v4] (d) {\(4\)};
    \vertex [left = 0.3cm of v4] (b) {\(6\)};
    \vertex [below =0.3cm of v4] (c) {\(5\)};
    \diagram* {
      (a) -- [double,line width=0.3mm] (v1) -- [double,line width=0.3mm] (e),
      (v1) -- [double,line width=0.3mm] (v4) -- [solid,line width=0.5mm] (c),
      (v4) -- [solid,line width=0.5mm] (b),
      (v4) -- [double,line width=0.3mm] (d)
    };
  \end{feynman}
\end{tikzpicture}}}
+
\raisebox{-3 mm}{\scalebox{0.7}{ 
  \begin{tikzpicture}
  \begin{feynman}[small]
    \vertex (a) {\(1\)};
    \vertex [right=of a] (v1);
    \vertex [right= 0.5cm of v1] (e) {\(3\)} ;
    \vertex [below = 0.5cm of v1] (v4);
    \vertex [right = 0.3cm of v4] (d) {\(4\)};
    \vertex [left = 0.3cm of v4] (b) {\(6\)};
    \vertex [below =0.3cm of v4] (c) {\(5\)};
    \diagram* {
      (a) -- [double,line width=0.3mm] (v1) -- [double,line width=0.3mm] (e),
      (v1) -- [double,line width=0.3mm] (v4) -- [double,line width=0.3mm] (c),
      (v4) -- [solid,line width=0.5mm] (b),
      (v4) -- [solid,line width=0.5mm] (d)
    };
  \end{feynman}
\end{tikzpicture}}}
+
\raisebox{-3 mm}{\scalebox{0.7}{ 
  \begin{tikzpicture}
  \begin{feynman}[small]
    \vertex (a) {\(1\)};
    \vertex [right=of a] (v1);
    \vertex [right= 0.5cm of v1] (e) {\(3\)} ;
    \vertex [below = 0.5cm of v1] (v4);
    \vertex [right = 0.3cm of v4] (d) {\(4\)};
    \vertex [left = 0.3cm of v4] (b) {\(6\)};
    \vertex [below =0.3cm of v4] (c) {\(5\)};
    \diagram* {
      (a) -- [double,line width=0.3mm] (v1) -- [double,line width=0.3mm] (e),
      (v1) -- [double,line width=0.3mm] (v4) -- [solid,line width=0.5mm] (c),
      (v4) -- [double,line width=0.3mm] (b),
      (v4) -- [solid,line width=0.5mm] (d)
    };
  \end{feynman}
\end{tikzpicture}}}
\end{aligned}
\right)
\end{aligned}\notag
    \end{small}
\end{equation}
as the diagrammatic equivalent of~\eqref{eq: 6-pt ac all}. This is an exact decomposition of the 6-pt amplitude due to our special choice~\eqref{fig:pqr} of the extended amplitudes; a similar expression appeared in~\cite{Cachazo:2016njl} with additional terms.

The factorisation for the $(1\to3)$ splitting follows the same pattern even at higher multiplicies,
\begin{equation}
\begin{split}
    A_n^{(1\to 3)}=& -\mathcal{A}_3^\text{ext}(1^{\Phi},2^{\Phi},3^{\Phi})\\
    &\times\sum_{i=4,\dotsc,n} s_{2i} \mathcal{A}_{n-1}^\text{ext} (3^{\Phi},\dots,i^{\Phi},\dots,1^{\Phi})\,.
\end{split}
\end{equation}
This is seen from diagram-by-diagram identification between contributions to the $n$-pt amplitude $\smash{A_n^{(1\to 3)}}$ and those to the $(n-1)$-pt extended amplitude $\mathcal{A}_{n-1}^\text{ext}$, obtained by erasing leaf leg $2$. We have checked the correspondence for $n=8$, and for a representative subset of the 88 diagrams contributing to the 10-pt amplitude.

Other splittings than $(1\to3)$ are more subtle to deal with. The simplest example, $\smash{A_{6}^{(1\to 4)}}$, can be written as \\[-4ex] 
\begin{multline}
    A_{6}^{(1\to 4)} - 
    \raisebox{-8 mm}{\scalebox{1}{\includegraphics[trim={0cm 0.2cm 0cm 0.2cm},clip,width=0.18\textwidth]{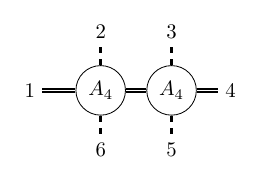}}}\\
    = -\sum_{\substack{i=2,3\\j=5,6}} s_{ij} \Tilde{A}_4^{i}(1,2,3,4)\Tilde{A}_4^{j}(4,5,6,1) \,,
    \label{eq: 6-pt ad all}
\end{multline}
with specific expressions for the 4-point amplitudes $\Tilde{A}_4^i$ (omitted for the sake of brevity). The main novelty is the presence of an exchange term on the left-hand side, expressed in terms of the \rev{partially off-shell 4-pt amplitudes $A_{4}$. These are now \emph{defined} by the indicated splitting, and thus given by~\eqref{A4-fact}}. Any such an exchange contribution is necessarily at least quadratic in the locus variables.

Another example is provided by the $(1\to5)$ splitting of the 8-pt amplitude, with diagrams listed in the Supplemental Material~\cite{SM}. We again find that the amplitude factorises modulo higher-order exchange diagrams,
\begin{align}
\notag
A_{8}^{(1\to 5)} - \text{exch} = 
&-\sum_{\substack{i=2,3,4\\j=6,7,8}}s_{ij}
     \mathcal{A}_5^\text{ext}(1^{\Phi},\dots,i^{\Phi},\dots,5^{\Phi})\\
     & \times\mathcal{A}_5^\text{ext}(5^{\Phi},\dots,j^{\Phi},\dots,1^{\Phi})\,,
\end{align}
where ``exch'' stands for
\begin{equation}
    \begin{aligned}
 & \raisebox{-9 mm}{\scalebox{0.7}{ 
  \begin{tikzpicture}
  \begin{feynman}[small]
    \vertex (a) {\(1\)};
   \node (v1) [circle, draw, right= 1 cm of a]{$A_{4}$};
   \node (v2) [circle, draw, right= 1.5 cm of v1]{$A_{6}$};
    \vertex [above = 1.2cm of v1] (b) {\(2\)};
    \vertex [below = 1.2cm of v1] (c) {\(8\)};
    \vertex [right = 1.2cm of v2] (h) {\(5\)};
    \vertex at ($(v2)!1!-120:(h)$) (g) {\(7\)};
    \vertex at ($(v2)!1!60:(h)$) (f) {\(4\)}; 
    \vertex at ($(v2)!1!120:(h)$) (e) {\(3\)};
    \vertex at ($(v2)!1!-60:(h)$) (d) {\(6\)};
    \diagram* {
      (a) -- [double,line width=0.3mm] (v1) -- [double,line width=0.3mm] (v2) --[dashed,line width=0.5mm] (d),
      (v2) -- [dashed,line width=0.5mm] (g),
      (v2) -- [double,line width=0.3mm] (h),
      (v2) -- [dashed,line width=0.5mm] (e),
      (v2) -- [dashed,line width=0.5mm] (f),
      (v1) -- [dashed,line width=0.5mm] (b),
      (v1) -- [dashed,line width=0.5mm] (c)
    };
  \end{feynman}
\end{tikzpicture}}}
+  
\raisebox{-9 mm}{\scalebox{0.7}{ 
  \begin{tikzpicture}
  \begin{feynman}[small]
    \vertex (a) {\(4\)};
   \node (v1) [circle, draw, below= 1.2 cm of a]{$A_{4}$};
   \node (v2) [circle, draw, left= 1.5 cm of v1]{$A_{6}$};
    \vertex [right = 1.2cm of v1] (b) {\(5\)};
    \vertex [below = 1.2cm of v1] (c) {\(6\)};
    \vertex [left = 1.2cm of v2] (h) {\(1\)};
    \vertex at ($(v2)!1!-120:(h)$) (g) {\(3\)};
    \vertex at ($(v2)!1!60:(h)$) (f) {\(8\)};
    \vertex at ($(v2)!1!120:(h)$) (e) {\(7\)};
    \vertex at ($(v2)!1!-60:(h)$) (d) {\(2\)};
    \diagram* {
      (a) -- [dashed,line width=0.5mm] (v1) -- [double,line width=0.3mm] (v2) --[dashed,line width=0.5mm] (d),
      (v2) -- [dashed,line width=0.5mm] (g),
      (v2) -- [double,line width=0.3mm] (h),
      (v2) -- [dashed,line width=0.5mm] (e),
      (v2) -- [dashed,line width=0.5mm] (f),
      (v1) -- [double,line width=0.3mm] (b),
      (v1) -- [dashed,line width=0.5mm] (c)
    };
  \end{feynman}
\end{tikzpicture}}}
\notag\\
 &\quad\quad- \raisebox{-9 mm}{\scalebox{0.7}{ 
  \begin{tikzpicture}
  \begin{feynman}[small]
    \vertex (a) {\(3\)};
   \node (v1) [circle, draw, below = 1.2 cm of a]{$A_{4}$};
   \node (v2) [circle, draw, left= 1.5 cm of v1]{$A_{4}$};
   \node (v3) [circle, draw, right = 1.5 cm of v1]{$A_{4}$};
    \vertex [below = 1.2cm of v1] (b) {\(7\)};
    \vertex [below = 1.2cm of v2] (c) {\(8\)};
    \vertex [left = 1.2cm of v2] (h) {\(1\)};
    \vertex [above= 1.2cm of v2] (g) {\(2\)};
    \vertex [above = 1.2cm of v3] (f) {\(4\)};
    \vertex [right = 1.2cm of v3] (e) {\(5\)};
    \vertex [below = 1.2cm of v3] (d) {\(6\)};
    \diagram* {
      (a) -- [dashed,line width=0.5mm] (v1),
      (v1) -- [dashed,line width=0.5mm] (b),
      (v1) -- [double,line width=0.3mm] (v3) --[dashed,line width=0.5mm] (d),
      (v1) -- [double,line width=0.3mm] (v2) -- [dashed,line width=0.5mm] (g),
      (v2) -- [double,line width=0.3mm] (h),
      (v3) -- [double,line width=0.3mm] (e),
      (v3) -- [dashed,line width=0.5mm] (f),
      (v2) -- [dashed,line width=0.5mm] (c)
    };
  \end{feynman}
\end{tikzpicture}}}\,,
\label{eq: 8-pt ae higher order}
\end{aligned} 
\end{equation}
\rev{defined in terms of partially off-shell amplitudes $A_4^{(1\to3)}$ and $A_6^{(1\to4)}$, given by \eqref{A4-fact} and \eqref{eq: 6-pt ad all}.} A similar expansion applies to $\smash{A_{8}^{(1\to 4)}}$, with a right-hand side factorised into even subamplitudes $\Tilde{A}^i_n$.

The examples we have worked out explicitly naturally lead to the following conjecture for all multiplicities. The NLSM partial amplitude $A_n$ for any odd splitting $(1\to k)$ can be represented exactly as  
\begin{align}
    \notag
    A_{n}^{(1\to k)}-\text{exch}=&-\sum_{\substack{1<i<k\\ k<j\leq n}} s_{ij}\mathcal{A}^\text{ext}_k(1^\Phi,\dots,i^\Phi,\dots, k^\Phi)\notag\\
    &\times\mathcal{A}^\text{ext}_{n-k+2}(k^\Phi,\dots,j^\Phi,\dots,1^\Phi)\,.
    \label{exact-factorisation-conjecture}
\end{align}
The precise form of the exchange terms at higher multiplicities remains an interesting open question. For even splittings, the amplitude factorises modulo exchange terms into an analogous expression with even subamplitudes $\Tilde{A}^i_n$. However, a general off-shell definition (generalising the on-shell results of \cite{Arkani-Hamed:2023swr}) of the latter does not appear as straightforward as in the case of odd splitting.



\emph{Discussion and Outlook}---In this Letter, we have introduced a new formulation of the NLSM, based on a set of covariant second-order field equations. This gives rise to a diagrammatic representation of tree-level scattering amplitudes that manifests the recently discovered hidden zero structure of the amplitudes~\cite{Arkani-Hamed:2023swr}. Our framework also naturally leads to a conjecture for an exact expansion of partial amplitudes in terms of the kinematic locus variables, whose leading, linear part manifests the near-zero factorisation of partial amplitudes into a product of two lower-point amplitudes of an extended theory. 

Operationally, our main tool was the choice of splitting $(1\to k)$ of the partial amplitude. This can be used to highlight the properties of amplitudes in specific regions of the kinematic space. Thus, for different $k$, our conjecture~\eqref{exact-factorisation-conjecture} gives different decompositions of the \emph{same} amplitude that manifest its asymptotic behavior near the respective kinematic loci~\eqref{eq: zero locus}. \del{In the future, it would be desirable to construct a detailed proof of our conjecture for odd splittings, and to understand better the decomposition of amplitudes for even splittings.}

We note that our formulation of the NLSM is similar in spirit to that of~\cite{Cheung:2021zvb}, which is based on the EoM for the adjoint flavour current of the NLSM. The latter is related by an off-shell mapping to the biadjoint scalar theory. \del{which naturally appears next to the Yang-Mills theory and general relativity under colour-kinematic duality~\cite{Bern:2008qj,Bern:2010ue}. (See also~\cite{Bern:2019prr} for a recent review.)} Our setup is instead related by an off-shell mapping to the Yang-Mills-scalar (YMS) theory~\cite{Cachazo:2014xea}. This follows from~\eqref{eomquartic}, whose structure is mathematically identical to the field equations of the YMS theory.

Our work opens several possible avenues for future research. First, the manifestly geometric structure of the field equations~\eqref{eomquartic} underlines the position of the NLSM in a web of exceptional scalar effective theories, including the (multiflavor) Dirac-Born-Infeld (DBI) theory and the special Galileon (sGal) theory. Unlike the NLSM, whose interactions include the exchange of a composite gluon by virtue of its mapping to the YMS theory, these other theories include coupling of the scalar matter to a composite graviton. Using the classical EoM, the relationship between the YMS theory, the NLSM, the DBI theory and the sGal theory can be made precise at the level of both scattering amplitudes and exact classical solutions. \del{We will report details in forthcoming publications.}

\del{A second avenue might focus on amplitudes beyond tree level. It is known that the EoM can be employed to compute loop integrands via the perturbiner approach or Berends-Giele recursion (see, for instance, \cite{Gomez:2022dzk,Lee:2022aiu}). Given that our formulation of the NLSM naturally facilitates expansion with respect to the locus variables, it will be of particular interest to investigate the expansion of loop integrands, and to compare the results with those obtained from the stringy-integral-based approach described in~\cite{Arkani-Hamed:2024fyd}.}

\rev{Furthermore}, it is known that the YMS theory admits higher-derivative double-copy compatible corrections~\cite{Carrasco:2022sck,Bonnefoy:2023imz}. There are indications that YMS$+\alpha'$ and NLSM$+\alpha'$ are related at the amplitude level~\cite{Carrasco:2022sck,Dong:2024klq}. It would be interesting to explore the possibility of a mapping between YMS$+\alpha'$ and NLSM$+\alpha'$ at the level of EoM, in the spirit of the present formulation.



\emph{Acknowledgements}---We thank Christoph Bartsch, Nichita Berzan, Song He and Jasper Roosmale Nepveu for helpful discussions and feedback on our results. T.W.~is supported by the China Scholarship Council.

\bibliographystyle{apsrev4-1}
\bibliography{refs}


\onecolumngrid
\newpage
\section*{End matter}
\twocolumngrid

\emph{Appendix A: Feynman Rules from the Equation of Motion}---Expanding the covariant derivatives in~\eqref{eomquartic}, the equations can be cast in a form suitable for iteration,
\begin{align}
\notag
        \Box \mc^{a}_{\mu}={}&2 f^{a}_{\alpha b}A^{\alpha}_{\nu}\partial^{\nu}\mc^{b}_{\mu} - f^{a}_{\alpha c}f^{c}_{\beta b}A^{\alpha\nu}A^{\beta}_{\nu}\mc^{b}_{\mu}- f^{a}_{\alpha b}f^{\alpha}_{cd}\mc^{b}_{\nu}\mc^{c\nu}\mc^{d}_{\mu}\,,\\
\notag
        \Box A^{\alpha}_{\mu}={}& f^\alpha_{ab} \mc^{a\nu} \partial_\mu \mc^b_\nu + f^\alpha_{\beta\gamma} \left(2A^\beta_{\nu}\partial^\nu A^\gamma_{\mu}- A^\beta_{\nu}\partial_\mu A^{\gamma \nu}\right) \\
        & - f^\alpha_{\beta\ve} f^\ve_{\gamma\delta} A^\beta_\nu A^{\gamma \nu} A^\delta_\mu - f^\alpha_{ac} f^c_{\beta b} A^{\beta}_{\mu}\mc^{a\nu}  \mc^{b}_\nu \,,
\label{NLSM_eq}
\end{align}
where we used the Lorenz gauge, $\partial^\mu A^\alpha_\mu=0$, for simplicity. In momentum space, the d'Alembertian operator on the left-hand side translates into a $-1/p^2$ propagator for any internal line of a diagram. The interaction vertices, extracted from \eqref{NLSM_eq}, read
\begin{align}
\parbox{3cm}{\scalebox{1}{ 
  \begin{tikzpicture}
  \begin{feynman}
    \vertex (c);
    \vertex [below right = 1.0 cm of c] (al) {$\alpha\lambda$};
   \vertex [above right = 1.0 cm of c] (b) {$b\nu$};
   \vertex [below left = 0.4 cm and 0.8cm of b](p) {$p$};
   \vertex [left = 1.0 cm of c] (a) {$a\mu$};
    +
    \diagram* {
      (a) -- [ solid,line width=0.5mm] (c) -- [ solid,line width=0.5mm] (b),
      (al) -- [boson,line width=0.5mm] (c)
    };
  \end{feynman}
\end{tikzpicture}}}
& = - 2\imag \eta_{\mu\nu}f^a_{\alpha b}p_\lambda\,,\nonumber\\
\parbox{3.25cm}{\scalebox{1}{ 
  \begin{tikzpicture}
  \begin{feynman}
    \vertex (c);
    \vertex [above right = 1.0 cm of c] (al) {$\alpha\kappa$};
    \vertex [below right = 1.0 cm of c] (be) {$\beta\lambda$};
   \vertex [right = 1.0 cm of c] (b) {$b\nu$};
   \vertex [left = 1.0 cm of c] (a) {$a\mu$};
    +
    \diagram* {
      (a) -- [ solid,line width=0.5mm] (c) -- [ solid,line width=0.5mm] (b),
      (al) -- [boson,line width=0.5mm] (c),
      (be) -- [boson,line width=0.5mm] (c)
    };
  \end{feynman}
\end{tikzpicture}}}
& = - \eta_{\mu\nu}\eta_{\kappa\lambda}(f^a_{\alpha c}f^c_{\beta b}+f^a_{\beta c}f^c_{\alpha b})\,,\nonumber\\
\parbox{3.25cm}{\scalebox{1}{ 
  \begin{tikzpicture}
  \begin{feynman}
    \vertex (c);
    \vertex [above right = 1.0 cm of c] (al) {$b\nu$};
    \vertex [below right = 1.0 cm of c] (be) {$d\lambda$};
   \vertex [right = 1.0 cm of c] (b) {$c\kappa$};
   \vertex [left = 1.0 cm of c] (a) {$a\mu$};
    +
    \diagram* {
      (a) -- [ solid,line width=0.5mm] (c) -- [ solid,line width=0.5mm] (b),
      (al) -- [solid,line width=0.5mm] (c),
      (be) -- [solid,line width=0.5mm] (c)
    };
  \end{feynman}
\end{tikzpicture}}}
& =
\begin{aligned}[t]
&- f^a_{\alpha b}f^\alpha_{cd}(\eta_{\mu\lambda}\eta_{\nu\kappa}-\eta_{\mu\kappa}\eta_{\nu\lambda})\\
&- f^a_{\alpha c}f^\alpha_{db}(\eta_{\mu\nu}\eta_{\kappa\lambda}-\eta_{\mu\lambda}\eta_{\nu\kappa})\\
&- f^a_{\alpha d}f^\alpha_{bc}(\eta_{\mu\kappa}\eta_{\nu\lambda}-\eta_{\mu\nu}\eta_{\kappa\lambda})\,,
\end{aligned}\nonumber\\
\parbox{3cm}{\scalebox{1}{ 
  \begin{tikzpicture}
  \begin{feynman}
    \vertex (c);
    \vertex [below right = 1.0 cm of c] (al) {$b\lambda$};
    \vertex [above left =0.4cm and 0.8cm of al] {$q$};
   \vertex [above right = 1.0 cm of c] (b) {$a\nu$};
   \vertex [below left =0.4cm and 0.8cm of b] {$p$};
   \vertex [left = 1.0 cm of c] (a) {$\alpha \mu$};
    +
    \diagram* {
      (a) -- [ boson,line width=0.5mm] (c) -- [ solid,line width=0.5mm] (b),
      (al) -- [solid,line width=0.5mm] (c)
    };
  \end{feynman}
\end{tikzpicture}}}
& =+\imag f^\alpha_{ab}\eta_{\nu \lambda} (p-q)_{\mu}\,,\nonumber\\
\parbox{3cm}{\scalebox{1}{ 
  \begin{tikzpicture}
  \begin{feynman}
    \vertex (c);
    \vertex [below right = 1.0 cm of c] (al) {$\gamma\lambda$};
   \vertex [above right = 1.0 cm of c] (b) {$\beta\nu$};
   \vertex [above left =0.4cm and 0.8cm of al] {$q$};
    \vertex [below left =0.4cm and 0.8cm of b] {$p$};
   \vertex [left = 1.0 cm of c] (a) {$\alpha \mu$};
    +
    \diagram* {
      (a) -- [ boson,line width=0.5mm] (c) -- [ boson,line width=0.5mm] (b),
      (al) -- [boson,line width=0.5mm] (c)
    };
  \end{feynman}
\end{tikzpicture}}}
& =\begin{aligned}[t]
+\imag f^\alpha_{\beta\gamma}[&2(\eta_{\mu\nu}p_\lambda-\eta_{\mu\lambda}q_\nu)\\
&-\eta_{\nu\lambda} (p-q)_\mu ]\,,
\end{aligned}\nonumber\\
\parbox{3.25cm}{\scalebox{1}{ 
  \begin{tikzpicture}
  \begin{feynman}
    \vertex (c);
    \vertex [above right = 1.0 cm of c] (al) {$\beta\nu$};
    \vertex [below right = 1.0 cm of c] (be) {$\delta\lambda$};
   \vertex [right = 1.0 cm of c] (b) {$\gamma\kappa$};
   \vertex [left = 1.0 cm of c] (a) {$\alpha\mu$};
    +
    \diagram* {
      (a) -- [boson,line width=0.5mm] (c) -- [ boson,line width=0.5mm] (b),
      (al) -- [boson,line width=0.5mm] (c),
      (be) -- [boson,line width=0.5mm] (c)
    };
  \end{feynman}
\end{tikzpicture}}}\,
&=
\begin{aligned}[t]
&-f^\alpha_{\beta\varepsilon}f^\varepsilon_{\gamma\delta}(\eta_{\mu\lambda}\eta_{\nu\kappa}-\eta_{\mu\kappa}\eta_{\nu\lambda})\\
&-f^\alpha_{\gamma\varepsilon}f^\varepsilon_{\delta\beta}(\eta_{\mu\nu}\eta_{\kappa\lambda}-\eta_{\mu\lambda}\eta_{\nu\kappa})\\
&-f^\alpha_{\delta\varepsilon}f^\varepsilon_{\beta\gamma}(\eta_{\mu\kappa}\eta_{\nu\lambda}-\eta_{\mu\nu}\eta_{\kappa\lambda})\,,
\end{aligned}\nonumber\\
\parbox{3.25cm}{\scalebox{1}{ 
  \begin{tikzpicture}
  \begin{feynman}
    \vertex (c);
    \vertex [above right = 1.0 cm of c] (al) {$a\mu$};
    \vertex [below right = 1.0 cm of c] (be) {$b\nu$};
   \vertex [right = 1.0 cm of c] (b) {$\beta\lambda$};
   \vertex [left = 1.0 cm of c] (a) {$\alpha\kappa$};
    +
    \diagram* {
      (a) -- [boson,line width=0.5mm] (c) -- [boson,line width=0.5mm] (b),
      (al) -- [solid,line width=0.5mm] (c),
      (be) -- [solid,line width=0.5mm] (c)
    };
  \end{feynman}
\end{tikzpicture}}}
& = - \eta_{\mu\nu}\eta_{\kappa\lambda}(f^\alpha_{ac}f^c_{\beta b}+f^\alpha_{bc}f^c_{\beta a})\,.\nonumber
\end{align}
Solid lines represent the covariant field $\mc^a_\mu$, whereas wavy lines represent the composite gauge field $A^\alpha_\mu$. All momenta indicated with labels $p,q$ are oriented from the leaves to the root, i.e.~from right to left.



\emph{Appendix B: Absence of Odd $\delta_{1i}$ Terms in Effective Vertices}---In proving the existence of hidden zeros diagram by diagram for odd splittings, we used the observation that the $\delta_{13}$ term in~\eqref{V4} is the only contribution among all the effective vertices $V_n$ containing $\delta_{1i}$ with an odd $i$. To see how this comes about, recall that only 1$\pi$I diagrams contribute to the effective vertices, and thus in any $V_n$ with $n\geq6$, the root leg is directly connected to a leaf leg by a vertex of type
$\smash{\parbox{7mm}{\scalebox{0.3}{ 
  \begin{tikzpicture}
  \begin{feynman}
    \vertex (c);
    \vertex [below right = 1.0 cm of c] (al) {};
   \vertex [above right = 1.0 cm of c] (b) {};
   \vertex [below left = 0.4 cm and 0.8cm of b](p) {};
   \vertex [left = 1.0 cm of c] (a) {};
    +
    \diagram* {
      (a) -- [ solid,line width=1mm] (c) -- [ solid,line width=1mm] (b),
      (al) -- [boson,line width=1mm] (c)
    };
  \end{feynman}
\end{tikzpicture}}}}$ or of type
$\smash{\parbox{8mm}{\scalebox{0.3}{ 
  \begin{tikzpicture}
  \begin{feynman}
    \vertex (c);
    \vertex [above right = 1.0 cm of c] (al) {};
    \vertex [below right = 1.0 cm of c] (be) {};
   \vertex [right = 1.0 cm of c] (b) {};
   \vertex [left = 1.0 cm of c] (a) {};
    +
    \diagram* {
      (a) -- [ solid,line width=1mm] (c) -- [ solid,line width=1mm] (b),
      (al) -- [boson,line width=1mm] (c),
      (be) -- [boson,line width=1mm] (c)
    };
  \end{feynman}
\end{tikzpicture}}}}$. The former will give either $\delta_{12}$ or $\delta_{1n}$. The latter will always give $\delta_{1i}$ with even $i$, since each internal gauge field propagator eventually branches into an even number of leaf legs. Note that the above argument relies on the explicit form of Feynman rules listed in Appendix A, whereby the two relevant types of vertices both contain a factor of $\eta_{\mu\nu}$ that couples the indices of the two solid lines.



\emph{Appendix C: Hidden Zeros of Even Splittings}---In the main text, we illustrated the vanishing of partial amplitudes of the NLSM on the kinematic locus~\eqref{eq: zero locus} for even splittings $(1\to k)$ by a simple example. Here we outline the general argument.

To start with, consider a single effective vertex $V_n$ lying on the splitting route  doubled line in our diagrams) of a diagram. Recall from Appendix B that for $n\geq6$, the root of the vertex must be attached to it via $\smash{\parbox{7mm}{\scalebox{0.3}{ 
  \begin{tikzpicture}
  \begin{feynman}
    \vertex (c);
    \vertex [below right = 1.0 cm of c] (al) {};
   \vertex [above right = 1.0 cm of c] (b) {};
   \vertex [below left = 0.4 cm and 0.8cm of b](p) {};
   \vertex [left = 1.0 cm of c] (a) {};
    +
    \diagram* {
      (a) -- [ solid,line width=1mm] (c) -- [ solid,line width=1mm] (b),
      (al) -- [boson,line width=1mm] (c)
    };
  \end{feynman}
\end{tikzpicture}}}}$ or
$\smash{\parbox{8mm}{\scalebox{0.3}{ 
  \begin{tikzpicture}
  \begin{feynman}
    \vertex (c);
    \vertex [above right = 1.0 cm of c] (al) {};
    \vertex [below right = 1.0 cm of c] (be) {};
   \vertex [right = 1.0 cm of c] (b) {};
   \vertex [left = 1.0 cm of c] (a) {};
    +
    \diagram* {
      (a) -- [ solid,line width=1mm] (c) -- [ solid,line width=1mm] (b),
      (al) -- [boson,line width=1mm] (c),
      (be) -- [boson,line width=1mm] (c)
    };
  \end{feynman}
\end{tikzpicture}}}}$. The Feynman rule for the latter includes the factor $\eta_{\kappa\lambda}$, which eventually translates to a set of dot products of momenta above and below the splitting route. Such a dot product also results from the $\delta_{13}\delta_{24}$ part of the $V_4$ vertex~\eqref{V4}. This leads to the key observation that upon dressing with polarisation vectors, any diagram with an effective vertex traversed by the splitting route vanishes on the corresponding kinematic locus~\eqref{eq: zero locus} unless the splitting route enters and leaves the vertex via two adjacent legs; see~\eqref{eq: three (4,4) diagrams} for an example.

Most diagrams will contain at least one vertex that is traversed by the splitting via non-adjacent legs; these diagrams vanish separately on the kinematic locus. The remaining diagrams are those where all (other than  doubled) legs in each individual vertex lie either above or below the splitting route. These can be subdivided into groups of diagrams with similar topologies, only differing by the order of effective vertices along the splitting route. We will now demonstrate that the sum of diagrams within each such group vanishes on the kinematic locus.

The argument consists of two separate claims. First, the numerators of all diagrams within a given group coincide on the kinematic locus. \rev{Let us start with a simple example, illustrating} the general pattern on the $(1\to 4)$ splitting of the 8-pt partial amplitude. For the sake of illustration, we choose a single group of diagrams \rev{(out of the complete list of diagrams contributing to the 8-pt partial amplitude in this splitting, which we include in the Supplemental Material~\cite{SM})}:
\begin{align}
\notag
&\parbox{0.25\textwidth}{\includegraphics[width=0.25\textwidth]{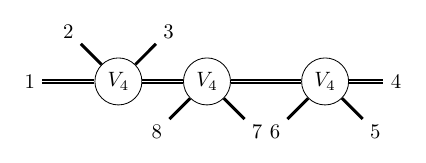}}
\parbox{0.25\textwidth}{\includegraphics[width=0.25\textwidth]{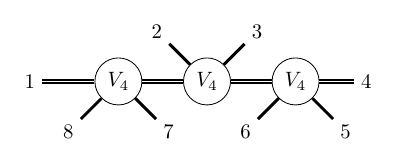}}\\
&\parbox{0.25\textwidth}{\includegraphics[width=0.25\textwidth]{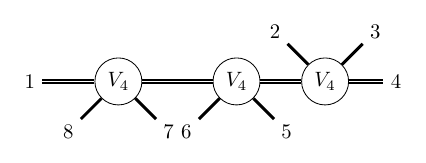}}
\label{eq: 8pt 1 4 3 diag}
\end{align}
Using~\eqref{V4}, the contribution of the $V_4$ vertex coupled to the two upper legs to the numerator of each of the three diagrams is proportional to
\begin{multline}
\varepsilon_2^{\mu_2}\varepsilon_3^{\mu_3}\left[ \raisebox{-5mm}{\scalebox{0.12}{\includegraphics[trim={0.3cm 0.3cm 0.3cm 0.2cm},clip,width=1\textwidth]{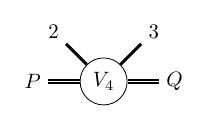}}}\right]\propto\eta_{\mu_P\mu_Q}\left(s_{23}+2Q\cdot (p_3-p_2)\right)\\
     \loceq \eta_{\mu_P\mu_Q}\left(s_{23}+2p_4\cdot(p_3-p_2)\right)\,.
     \label{eq: ever numerator same examp}
\end{multline}
For the diagrams in~\eqref{eq: ever numerator same examp}, the momenta $(P,Q)$ equal, respectively, $(p_1,p_4+\dotsb+p_8)$, $(p_7+p_8+p_1,p_4+p_5+p_6)$, and $(p_5+\dotsb+p_1,p_4)$. On the $(1\to4)$ locus, the numerator takes the same value for all three diagrams. Similar logic applies to the numerator contributions of the two $V_4$ vertices in~\eqref{eq: 8pt 1 4 3 diag} coupled to the four lower legs.

\rev{We now outline a general proof of the equivalence of numerators within a group of diagrams related by a permutation of effective vertices along the splitting route. Generalizing the notation in~\eqref{eq: ever numerator same examp}, we represent a single vertex on the splitting route by the following diagram (the argument for vertices whose all other than  doubled legs lie below the splitting route is analogous):
\begin{equation}
    \raisebox{-5mm}{\scalebox{0.15}{\includegraphics[trim={0.3cm 0.3cm 0.3cm 0.2cm},clip,width=1\textwidth]{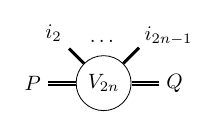}}}
\label{V2n}
\end{equation}
As the vertex is swapped with other vertices along the splitting route, the momenta $P$, $Q$ will change but the rest of the vertex (including any subdiagrams attached to it via the black legs) will remain the same. We want to show that the change in $P,Q$ will not affect the value of the vertex on the kinematic locus~\eqref{eq: zero locus}.}

\rev{It follows from the Feynman rules listed in Appendix A that the contribution of the $V_{2n}$ vertex is given by $\eta_{\mu_P\mu_Q}$ times a rational function of the momenta $p_{i_2},\dotsc, p_{i_{2n-1}}$ and $Q$. The denominator of this function does not contain $Q$; cf.~the expression for $V_6$ in~\eqref{V6}. Also, the $V_{2n}$ vertex is a linear function of $Q$, which eventually appears in a dot product with one of $p_{i_k}$ with $k=2,\dotsc,2n-1$, or with the momentum of another external leg, connected to $V_{2n}$ by a propagator on an $i_k$ leg. Finally, observe that as $V_{2n}$ is moved along the splitting route, $Q$ will only change by a linear combination of momenta below the splitting route. This guarantees that on the kinematic locus~\eqref{eq: zero locus}, the contribution of the vertex will not change. By a trivial extension of the argument to all vertices on the splitting route, we conclude that the numerators of diagrams differing by a mere permutation of vertices along the splitting route are equal on the locus.}

The second issue we have to deal with is that, even with equal numerators, the denominators of diagrams related by a permutation of effective vertices along the splitting route differ. However, within a single group, the (products of) propagators add up to zero on the kinematic locus. Let us first illustrate this on the diagrams shown in~\eqref{eq: 8pt 1 4 3 diag}. Here the propagators of the diagrams add up to
\begin{align}
\notag
   &\frac{1}{s_{123}s_{12378}}+ \frac{1}{s_{178}s_{12378}}+ \frac{1}{s_{178}s_{15678}}\\
   \notag
   &\loceq\frac{1}{s_{123}(s_{123}+s_{178})}+\frac{1}{s_{178}(s_{123}+s_{178})}+\frac{1}{s_{178}s_{15678}}\\
   &=\frac{1}{s_{178}}\left(\frac{1}{s_{123}}+\frac{1}{s_{15678}}\right)\,.
\end{align}
On the $(1\to 4)$ locus, the expression in the parentheses is proportional to $s_{123}+s_{15678}\loceq s_{1235678}=s_4=0$, in accord with the expectation.

In general, the clustering of the external legs above and below the splitting route into effective vertices defines an ordered partition $U\equiv\{\mfu_1,\mfu_2,\dots,\mfu_u\}$ of length $u$ of the set $\{2,3,\dots,k-1\}$, and a partition $D=\{\mfd_1,\mfd_2,\dots,\mfd_d\}$ of length $d$ of the set $\{n,n-1,\dotsc,k+1\}$. The set of all diagrams in a single group maps to the elements of the shuffle product $U\shuffle D$ such that the relative ordering within each of $U$ and $D$ is maintained. The sum of denominators of the diagrams in the group then corresponds to the left-hand side of the following identity,
\begin{equation}
\begin{small}
    \begin{gathered}
       \sum_{\sigma\in U\shuffle D}\prod_{r=1}^{u+d-1}\frac{1}{s_{1\bigcup_{w=1}^{r} \sigma_w}}
        \loceq\prod_{j=1}^{u-1}\frac{1}{s_{1\bigcup_{\ell=1}^{j} \mfu_{\ell}}}
        \phantom{large space}\\
        \phantom{space}\times
        \prod_{v=1}^{d-1}\frac{1}{s_{1\bigcup_{m=1}^{v} \mfd_m}}
        \Biggl(\frac{1}{s_{1\bigcup_{{\ell}=1}^{u} \mfu_{\ell}}}+\frac{1}{s_{1\bigcup_{m=1}^{d} \mfd_m}}\Biggr)\,,
        \label{eq: block shuffle identity}
    \end{gathered}
\end{small}
\end{equation}
where $s_{1\sigma_w}$ denotes the Mandelstam variable with labels $\{1\}\cup\sigma_w$, and $\sigma_w$ is the $w$-th element of $ \sigma \in U\shuffle D$. The expression in the parentheses on the right-hand side vanishes on the $(1\to k)$ locus, being proportional to $s_{1\bigcup_{{\ell}=1}^{u} \mfu_{\ell}}+s_{1\bigcup_{m=1}^{d} \mfd_m}\loceq s_{\{1\dotsb n\}\backslash\{k\}}=s_k=0$. This completes the proof of the vanishing of partial amplitudes of the NLSM on the kinematic locus~\eqref{eq: zero locus} with even $k$. 


\balancecolsandclearpage

\onecolumngrid

\section*{Supplemental Material}
\label{app: 8-pt diagrams}

\noindent \rev{\emph{Effective Vertices and Graphs}---The sets of 1$\pi$I flavour-ordered 4-, 6- and 8-pt graphs, that together define the effective vertices, are given by:
\vspace{1cm}
\begin{flalign*}\hspace{-6cm}
\raisebox{-10 mm}{\scalebox{0.9}{
\begin{tikzpicture}
  \begin{feynman}[small]
    \vertex (a) {\(1\)};
    \node (v1) [circle, draw, right= 1.2 cm of a]{$V_{4}$};
    \vertex [right = 1.2cm of v1] (c) {\(3\)};
    \vertex [below right =  1.4cm of v1] (d) {\(4\)};
    \vertex [above right = 1.4cm of v1] (b) {\(2\)};
    \diagram* {
      (a) -- [solid,line width=0.5mm] (v1) --[solid,line width=0.5mm] (b),
      (v1) -- [solid,line width=0.5mm] (d),
      (v1) -- [solid,line width=0.5mm] (c)
    };
  \end{feynman}
\end{tikzpicture}}}
\equiv
\raisebox{-8 mm}{\scalebox{0.9}{
\begin{tikzpicture}
  \begin{feynman}[small]
    \vertex (a) {\(1\)};
    \vertex (v1) [right= 0.8 cm of a];
    \vertex [right = 0.8cm of v1] (c) {\(3\)};
    \vertex [below right =  0.8cm of v1] (d) {\(4\)};
    \vertex [above right = 0.8cm of v1] (b) {\(2\)};
    \diagram* {
      (a) -- [solid,line width=0.5mm] (v1) --[solid,line width=0.5mm] (b),
      (v1) -- [solid,line width=0.5mm] (d),
      (v1) -- [solid,line width=0.5mm] (c)
    };
  \end{feynman}
\end{tikzpicture}}}
+
\raisebox{-8 mm}{\scalebox{0.9}{
\begin{tikzpicture}
  \begin{feynman}[small]
    \vertex (a) {\(1\)};
    \vertex (v1) [right= 0.8 cm of a];
    \vertex (v2) [below right= 0.8 cm of v1];
    \vertex [above right = 0.6 cm of v2] (c) {\(3\)};
    \vertex [below right =  0.6 cm of v2] (d) {\(4\)};
    \vertex [above right = 0.8 cm of v1] (b) {\(2\)};
    \diagram* {
      (a) -- [solid,line width=0.5mm] (v1)-- [boson,line width=0.5mm] (v2) --[solid,line width=0.5mm] (d),
      (v1) -- [solid,line width=0.5mm] (b),
      (v2) -- [solid,line width=0.5mm] (c)
    };
  \end{feynman}
\end{tikzpicture}}}
+\raisebox{-8 mm}{\scalebox{0.9}{
\begin{tikzpicture}
  \begin{feynman}[small]
    \vertex (a) {\(1\)};
    \vertex (v1) [right= 0.8 cm of a];
    \vertex (v2) [above right= 0.8 cm of v1];
    \vertex [above right = 0.6 cm of v2] (c) {\(2\)};
    \vertex [below right =  0.6 cm of v2] (d) {\(3\)};
    \vertex [below right = 0.8 cm of v1] (b) {\(4\)};
    \diagram* {
      (a) -- [solid,line width=0.5mm] (v1)-- [boson,line width=0.5mm] (v2) --[solid,line width=0.5mm] (d),
      (v1) -- [solid,line width=0.5mm] (b),
      (v2) -- [solid,line width=0.5mm] (c)
    };
  \end{feynman}
\end{tikzpicture}}},    
\end{flalign*}
\vspace{0.5cm}
\begin{flalign*}\hspace{-4.5cm}
\raisebox{-12 mm}{\scalebox{0.9}{
\begin{tikzpicture}
  \begin{feynman}[small]
    \vertex (a) {\(1\)};
    \node (v1) [circle, draw, right= 1.2 cm of a]{$V_{6}$};
    \vertex [right = 1.2cm of v1] (c) {\(4\)};
    \vertex [above = 1.2cm of v1] (e) {\(2\)};
    \vertex [below = 1.2cm of v1] (f) {\(6\)};
    \vertex [below right =  1.4cm of v1] (d) {\(5\)};
    \vertex [above right = 1.4cm of v1] (b) {\(3\)};
    \diagram* {
      (a) -- [solid,line width=0.5mm] (v1) --[solid,line width=0.5mm] (b),
      (v1) -- [solid,line width=0.5mm] (d),
      (v1) -- [solid,line width=0.5mm] (c),
      (v1) -- [solid,line width=0.5mm] (e),
      (v1) -- [solid,line width=0.5mm] (f),
    };
  \end{feynman}
\end{tikzpicture}}}
\equiv &
 \raisebox{-13 mm}{\scalebox{0.9}{
\begin{tikzpicture}
  \begin{feynman}[small]
    \vertex (a) {\(1\)};
    \vertex (v1) [right= 1 cm of a];
     \vertex [above right =0.8 cm of v1] (v2);
    \vertex [below right =0.8 cm of v1] (v3);
    \vertex [above right = 0.7cm of v2] (b) {\(2\)};
    \vertex [right = 0.7 cm of v2] (c) {\(3\)};
    \vertex [below right = 0.7cm of v3] (f) {\(6\)};
    \vertex [right = 0.7cm of v3] (e) {\(5\)};
    \vertex [right = 1.2 cm of v1] (d) {\(4\)};
    \diagram* {
      (a) -- [solid,line width=0.5mm] (v1) -- [solid,line width=0.5mm] (d),
      (v3) -- [boson,line width=0.5mm] (v1) -- [boson,line width=0.5mm] (v2),
      (v3) -- [solid,line width=0.5mm] (f),
      (v3) -- [solid,line width=0.5mm] (e),
      (v2) -- [solid,line width=0.5mm] (b),
      (v2) -- [solid,line width=0.5mm] (c)
    };
  \end{feynman}
\end{tikzpicture}}}
  \\
&
+\left(
\begin{aligned}
&\raisebox{-12 mm}{\scalebox{0.9}{
\begin{tikzpicture}
  \begin{feynman}[small]
    \vertex (a) {\(1\)};
    \vertex (v1) [right= 1 cm of a];
     \vertex [right =0.8 cm of v1] (v2);
    \vertex [below right =0.8 cm of v1] (v3);
    \vertex [above right = 0.7cm of v2] (b) {\(3\)};
    \vertex [right = 0.7 cm of v2] (c) {\(4\)};
    \vertex [below right = 0.7cm of v3] (f) {\(6\)};
    \vertex [right = 0.7cm of v3] (e) {\(5\)};
    \vertex [above right = 1 cm of v1] (d) {\(2\)};
    \diagram* {
      (a) -- [solid,line width=0.5mm] (v1) -- [solid,line width=0.5mm] (d),
      (v3) -- [boson,line width=0.5mm] (v1) -- [boson,line width=0.5mm] (v2),
      (v3) -- [solid,line width=0.5mm] (f),
      (v3) -- [solid,line width=0.5mm] (e),
      (v2) -- [solid,line width=0.5mm] (b),
      (v2) -- [solid,line width=0.5mm] (c)
    };
  \end{feynman}
\end{tikzpicture}}}
 +
\raisebox{-12 mm}{\scalebox{0.9}{
\begin{tikzpicture}
  \begin{feynman}[small]
    \vertex (a) {\(1\)};
    \vertex (v1) [right= 1 cm of a];
     \vertex [above right =0.8 cm of v1] (b) {\(2\)};
    \vertex [below right =0.5 cm of v1] (v2);
    \vertex [above right =0.5 cm of v2] (v3);
    \vertex [below right =0.5 cm of v2] (v4);
    \vertex [above right = 0.7cm of v3] (c) {\(3\)};
    \vertex [right = 0.7cm of v3] (d) {\(4\)};
    \vertex [right = 0.7cm of v4] (e) {\(5\)};
    \vertex [below right = 0.7cm of v4] (f) {\(6\)};
    \diagram* {
      (a) -- [solid,line width=0.5mm] (v1) -- [solid,line width=0.5mm] (b),
      (v1) -- [boson,line width=0.5mm] (v2) -- [boson,line width=0.5mm] (v3) -- [solid,line width=0.5mm] (c),
      (v3) -- [solid,line width=0.5mm] (d),
      (v2) -- [boson,line width=0.5mm] (v4) -- [solid,line width=0.5mm] (e),
      (v4) -- [solid,line width=0.5mm] (f)
    };
  \end{feynman}
\end{tikzpicture}}}
+
\raisebox{-12 mm}{\scalebox{0.9}{
\begin{tikzpicture}
  \begin{feynman}[small]
    \vertex (a) {\(1\)};
    \vertex (v1) [right= 1 cm of a];
     \vertex [above right =0.8 cm of v1] (b) {\(2\)};
    \vertex [below right =0.8 cm of v1] (v2);
    \vertex [right =0.8 cm of v2] (v3);
    \vertex [above right =0.8 cm of v2] (c) {\(3\)};
    \vertex [below right =0.8 cm of v2] (f) {\(6\)};
    \vertex [above right = 0.7cm of v3] (d) {\(4\)};
    \vertex [below right = 0.7cm of v3] (e) {\(5\)};
    \diagram* {
      (a) -- [solid,line width=0.5mm] (v1) -- [solid,line width=0.5mm] (b),
      (v1) -- [boson,line width=0.5mm] (v2) -- [boson,line width=0.5mm] (v3) -- [solid,line width=0.5mm] (d),
      (v2) -- [solid,line width=0.5mm] (c),
      (v2) -- [solid,line width=0.5mm] (f),
      (v3) -- [solid,line width=0.5mm] (e)
    };
  \end{feynman}
\end{tikzpicture}}}\\
&+
\raisebox{-12 mm}{\scalebox{0.9}{
\begin{tikzpicture}
  \begin{feynman}[small]
    \vertex (a) {\(1\)};
    \vertex (v1) [right= 1 cm of a];
     \vertex [above right =0.8 cm of v1] (b) {\(2\)};
    \vertex [below right =0.5 cm of v1] (v2);
    \vertex [below right =0.8 cm of v2] (v3);
    \vertex [above right =0.8 cm of v2] (c) {\(3\)};
    \vertex [right =0.8 cm of v2] (f) {\(4\)};
    \vertex [right = 0.7cm of v3] (d) {\(5\)};
    \vertex [below right = 0.7cm of v3] (e) {\(6\)};
    \diagram* {
      (a) -- [solid,line width=0.5mm] (v1) -- [solid,line width=0.5mm] (b),
      (v1) -- [boson,line width=0.5mm] (v2) -- [boson,line width=0.5mm] (v3) -- [solid,line width=0.5mm] (d),
      (v2) -- [solid,line width=0.5mm] (c),
      (v2) -- [solid,line width=0.5mm] (f),
      (v3) -- [solid,line width=0.5mm] (e)
    };
  \end{feynman}
\end{tikzpicture}}}
+
\raisebox{-12 mm}{\scalebox{0.9}{
\begin{tikzpicture}
  \begin{feynman}[small]
    \vertex (a) {\(1\)};
    \vertex (v1) [right= 1 cm of a];
     \vertex [above right =0.8 cm of v1] (b) {\(2\)};
    \vertex [below right =0.8 cm of v1] (v2);
    \vertex [above right =0.8 cm of v2] (v3);
    \vertex [right =0.8 cm of v2] (c) {\(5\)};
    \vertex [below right =0.8 cm of v2] (f) {\(6\)};
    \vertex [above right = 0.7cm of v3] (d) {\(3\)};
    \vertex [right = 0.7cm of v3] (e) {\(4\)};
    \diagram* {
      (a) -- [solid,line width=0.5mm] (v1) -- [solid,line width=0.5mm] (b),
      (v1) -- [boson,line width=0.5mm] (v2) -- [boson,line width=0.5mm] (v3) -- [solid,line width=0.5mm] (d),
      (v2) -- [solid,line width=0.5mm] (c),
      (v2) -- [solid,line width=0.5mm] (f),
      (v3) -- [solid,line width=0.5mm] (e)
    };
  \end{feynman}
\end{tikzpicture}}}\,\,\, + \,\,\, \text{mirror}
\end{aligned}
\right),\notag
\end{flalign*}}
\vspace{0.5cm}
\begin{align*}
\raisebox{-12 mm}{\scalebox{0.9}{
\begin{tikzpicture}
  \begin{feynman}[small]
    \vertex (a) {\(1\)};
    \node (v1) [circle, draw, right= 1.2 cm of a]{$V_{8}$};
    \vertex [right = 1.2cm of v1] (c) {\(5\)};
    \vertex [above = 1.2cm of v1] (e) {\(2\)};
    \vertex [below = 1.2cm of v1] (f) {\(8\)};
    \vertex [below right =  0.6 cm and 1.1 cm of v1] (d) {\(6\)};
    \vertex [below right =  1.1 cm and 0.6 cm of v1] (g) {\(7\)};
    \vertex [above right = 0.6 cm and 1.1 cm of v1] (b) {\(4\)};
    \vertex [above right = 1.1 cm and 0.6 cm of v1] (h) {\(3\)};
    \diagram* {
      (a) -- [solid,line width=0.5mm] (v1) --[solid,line width=0.5mm] (b),
      (v1) -- [solid,line width=0.5mm] (d),
      (v1) -- [solid,line width=0.5mm] (c),
      (v1) -- [solid,line width=0.5mm] (e),
      (v1) -- [solid,line width=0.5mm] (f),
      (v1) -- [solid,line width=0.5mm] (g),
      (v1) -- [solid,line width=0.5mm] (h),
    };
  \end{feynman}
\end{tikzpicture}}}
\equiv
&
\raisebox{-10 mm}{\scalebox{0.7}{
\begin{tikzpicture}
  \begin{feynman}[small]
    \vertex (a) {\(1\)};
    \vertex (v1) [right= 1 cm of a];
    \vertex [below right =0.8 cm of v1] (v2);
    \vertex [above right =0.8 cm of v2] (v3);
    \vertex [below right =0.8 cm of v2] (v4);
    \vertex [right =0.6 cm of v4] (v5);
    \vertex [below right =0.8 cm of v4] (v6);
    \vertex [above right = 0.9 cm of v1] (b) {\(2\)};
    \vertex [above right = 0.7 cm of v3] (c) {\(3\)};
    \vertex [right = 0.7 cm of v3] (d) {\(4\)};
    \vertex [above right = 0.7 cm of v5] (e) {\(5\)};
    \vertex [right = 0.7 cm of v5] (f) {\(6\)};
   \vertex [right = 0.7 cm of v6] (g) {\(7\)};
    \vertex [below right = 0.7 cm of v6] (h) {\(8\)};
    \diagram* {
      (a) -- [solid,line width=0.5mm] (v1) -- [solid,line width=0.5mm] (b),
      (v1) -- [boson,line width=0.5mm] (v2) -- [boson,line width=0.5mm] (v3),
      (v2) -- [boson,line width=0.5mm] (v4) -- [boson,line width=0.5mm] (v5),
      (v4) -- [boson,line width=0.5mm] (v6),
      (v3) -- [solid,line width=0.5mm] (c),
      (v3) -- [solid,line width=0.5mm] (d),
      (v5) -- [solid,line width=0.5mm] (e),
      (v5) -- [solid,line width=0.5mm] (f),
      (v6) -- [solid,line width=0.5mm] (g),
      (v6) -- [solid,line width=0.5mm] (h)
    };
  \end{feynman}
\end{tikzpicture}}}
+
\raisebox{-10 mm}{\scalebox{0.7}{
\begin{tikzpicture}
  \begin{feynman}[small]
    \vertex (a) {\(1\)};
    \vertex (v1) [right= 1 cm of a];
    \vertex [below right =0.8 cm of v1] (v2);
    \vertex [above right =0.8 cm of v2] (v3);
    \vertex [below right =0.8 cm of v2] (v4);
    \vertex [below right =0.8 cm of v4] (v5);
    \vertex [above right = 0.9 cm of v1] (b) {\(2\)};
    \vertex [above right = 0.7 cm of v4] (e) {\(5\)};
    \vertex [right = 0.7 cm of v5] (g) {\(7\)};
    \vertex [below right = 0.7 cm of v5] (h) {\(8\)};
    \vertex [right = 0.7 cm of v4] (f) {\(6\)};
    \vertex [above right = 0.7 cm of v3] (c) {\(3\)};
    \vertex [right = 0.7 cm of v3] (d) {\(4\)};
    \diagram* {
      (a) -- [solid,line width=0.5mm] (v1) -- [solid,line width=0.5mm] (b),
      (v1) -- [boson,line width=0.5mm] (v2) -- [boson,line width=0.5mm] (v3),
      (v2) -- [boson,line width=0.5mm] (v4) -- [boson,line width=0.5mm] (v5),
      (v4) -- [solid,line width=0.5mm] (e),
      (v5) -- [solid,line width=0.5mm] (g),
      (v5) -- [solid,line width=0.5mm] (h),
      (v4) -- [solid,line width=0.5mm] (f),
      (v3) -- [solid,line width=0.5mm] (c),
      (v3) -- [solid,line width=0.5mm] (d)
    };
  \end{feynman}
\end{tikzpicture}}}
+
\raisebox{-10 mm}{\scalebox{0.7}{
\begin{tikzpicture}
  \begin{feynman}[small]
    \vertex (a) {\(1\)};
    \vertex (v1) [right= 1 cm of a];
    \vertex [below right =0.8 cm of v1] (v2);
    \vertex [above right =0.8 cm of v2] (v3);
    \vertex [below right =0.8 cm of v2] (v4);
    \vertex [right =0.8 cm of v4] (v5);
    \vertex [above right = 0.9 cm of v1] (b) {\(2\)};
    \vertex [above right = 0.7 cm of v4] (e) {\(5\)};
    \vertex [above right = 0.7 cm of v5] (f) {\(6\)};
    \vertex [below right = 0.7 cm of v5] (g) {\(7\)};
    \vertex [below right = 0.7 cm of v4] (h) {\(8\)};
    \vertex [above right = 0.7 cm of v3] (c) {\(3\)};
    \vertex [right = 0.7 cm of v3] (d) {\(4\)};
    \diagram* {
      (a) -- [solid,line width=0.5mm] (v1) -- [solid,line width=0.5mm] (b),
      (v1) -- [boson,line width=0.5mm] (v2) -- [boson,line width=0.5mm] (v3),
      (v2) -- [boson,line width=0.5mm] (v4) -- [boson,line width=0.5mm] (v5),
      (v4) -- [solid,line width=0.5mm] (e),
      (v5) -- [solid,line width=0.5mm] (f),
      (v5) -- [solid,line width=0.5mm] (g),
      (v4) -- [solid,line width=0.5mm] (h),
      (v3) -- [solid,line width=0.5mm] (c),
      (v3) -- [solid,line width=0.5mm] (d)
    };
  \end{feynman}
\end{tikzpicture}}}
+
\raisebox{-10 mm}{\scalebox{0.7}{
\begin{tikzpicture}
  \begin{feynman}[small]
    \vertex (a) {\(1\)};
    \vertex (v1) [right= 1 cm of a];
    \vertex [below right =0.8 cm of v1] (v2);
    \vertex [above right =0.8 cm of v2] (v3);
    \vertex [below right =0.8 cm of v2] (v4);
    \vertex [above right =0.8 cm of v4] (v5);
    \vertex [above right = 0.9 cm of v1] (b) {\(2\)};
    \vertex [right = 0.7 cm of v4] (g) {\(7\)};
    \vertex [above right = 0.7 cm of v5] (e) {\(5\)};
    \vertex [right = 0.7 cm of v5] (f) {\(6\)};
    \vertex [below right = 0.7 cm of v4] (h) {\(8\)};
    \vertex [above right = 0.7 cm of v3] (c) {\(3\)};
    \vertex [right = 0.5 cm of v3] (d) {\(4\)};
    \diagram* {
      (a) -- [solid,line width=0.5mm] (v1) -- [solid,line width=0.5mm] (b),
      (v1) -- [boson,line width=0.5mm] (v2) -- [boson,line width=0.5mm] (v3),
      (v2) -- [boson,line width=0.5mm] (v4) -- [boson,line width=0.5mm] (v5),
      (v5) -- [solid,line width=0.5mm] (e),
      (v4) -- [solid,line width=0.5mm] (g),
      (v4) -- [solid,line width=0.5mm] (h),
      (v5) -- [solid,line width=0.5mm] (f),
      (v3) -- [solid,line width=0.5mm] (c),
      (v3) -- [solid,line width=0.5mm] (d)
    };
  \end{feynman}
\end{tikzpicture}}}\\
&
+
\raisebox{-10 mm}{\scalebox{0.7}{
\begin{tikzpicture}
  \begin{feynman}[small]
    \vertex (a) {\(1\)};
    \vertex (v1) [right= 1 cm of a];
    \vertex [below right =0.8 cm of v1] (v2);
    \vertex [below right =0.8 cm of v2] (v3);
    \vertex [above right =0.8 cm of v2] (v4);
    \vertex [above right =0.8 cm of v4] (v5);
    \vertex [right =0.6 cm of v4] (v6);
    \vertex [above right = 0.9 cm of v1] (b) {\(2\)};
    \vertex [right = 0.7 cm of v3] (g) {\(7\)};
    \vertex [below right = 0.7 cm of v3] (h) {\(8\)};
    \vertex [right = 0.7 cm of v5] (d) {\(4\)};
    \vertex [above right = 0.7 cm of v5] (c) {\(3\)};
   \vertex [right = 0.7 cm of v6] (e) {\(5\)};
    \vertex [below right = 0.7 cm of v6] (f) {\(6\)};
    \diagram* {
      (a) -- [solid,line width=0.5mm] (v1) -- [solid,line width=0.5mm] (b),
      (v1) -- [boson,line width=0.5mm] (v2) -- [boson,line width=0.5mm] (v3),
      (v2) -- [boson,line width=0.5mm] (v4) -- [boson,line width=0.5mm] (v5),
      (v4) -- [boson,line width=0.5mm] (v6),
      (v3) -- [solid,line width=0.5mm] (g),
      (v3) -- [solid,line width=0.5mm] (h),
      (v5) -- [solid,line width=0.5mm] (c),
      (v5) -- [solid,line width=0.5mm] (d),
      (v6) -- [solid,line width=0.5mm] (e),
      (v6) -- [solid,line width=0.5mm] (f)
    };
  \end{feynman}
\end{tikzpicture}}}
+
\raisebox{-10 mm}{\scalebox{0.7}{
\begin{tikzpicture}
  \begin{feynman}[small]
    \vertex (a) {\(1\)};
    \vertex (v1) [right= 1 cm of a];
    \vertex [below right =0.8 cm of v1] (v2);
    \vertex [below right =0.8 cm of v2] (v3);
    \vertex [above right =0.8 cm of v2] (v4);
    \vertex [below right =0.8 cm of v4] (v5);
    \vertex [above right = 0.9 cm of v1] (b) {\(2\)};
    \vertex [above right = 0.7 cm of v4] (c) {\(3\)};
    \vertex [right = 0.5 cm of v4] (d) {\(4\)};
    \vertex [above right = 0.7 cm of v5] (e) {\(5\)};
    \vertex [right = 0.7 cm of v5] (f) {\(6\)};
    \vertex [right = 0.7 cm of v3] (g) {\(7\)};
    \vertex [below right = 0.7 cm of v3] (h) {\(8\)};
    \diagram* {
      (a) -- [solid,line width=0.5mm] (v1) -- [solid,line width=0.5mm] (b),
      (v1) -- [boson,line width=0.5mm] (v2) -- [boson,line width=0.5mm] (v3),
      (v2) -- [boson,line width=0.5mm] (v4) -- [boson,line width=0.5mm] (v5),
      (v4) -- [solid,line width=0.5mm] (c),
      (v4) -- [solid,line width=0.5mm] (d),
      (v5) -- [solid,line width=0.5mm] (e),
      (v5) -- [solid,line width=0.5mm] (f),
      (v3) -- [solid,line width=0.5mm] (g),
      (v3) -- [solid,line width=0.5mm] (h)
    };
  \end{feynman}
\end{tikzpicture}}}
+
\raisebox{-10 mm}{\scalebox{0.7}{
\begin{tikzpicture}
  \begin{feynman}[small]
    \vertex (a) {\(1\)};
    \vertex (v1) [right= 1 cm of a];
    \vertex [below right =0.8 cm of v1] (v2);
    \vertex [below right =0.8 cm of v2] (v3);
    \vertex [above right =0.8 cm of v2] (v4);
    \vertex [right =0.8 cm of v4] (v5);
    \vertex [above right = 0.9 cm of v1] (b) {\(2\)};
    \vertex [above right = 0.7 cm of v4] (c) {\(3\)};
    \vertex [above right = 0.7 cm of v5] (d) {\(4\)};
    \vertex [below right = 0.7 cm of v5] (e) {\(5\)};
    \vertex [below right = 0.7 cm of v4] (f) {\(6\)};
    \vertex [right = 0.7 cm of v3] (g) {\(7\)};
    \vertex [below right = 0.7 cm of v3] (h) {\(8\)};
    \diagram* {
      (a) -- [solid,line width=0.5mm] (v1) -- [solid,line width=0.5mm] (b),
      (v1) -- [boson,line width=0.5mm] (v2) -- [boson,line width=0.5mm] (v3),
      (v2) -- [boson,line width=0.5mm] (v4) -- [boson,line width=0.5mm] (v5),
      (v4) -- [solid,line width=0.5mm] (c),
      (v5) -- [solid,line width=0.5mm] (d),
      (v5) -- [solid,line width=0.5mm] (e),
      (v4) -- [solid,line width=0.5mm] (f),
      (v3) -- [solid,line width=0.5mm] (g),
      (v3) -- [solid,line width=0.5mm] (h)
    };
  \end{feynman}
\end{tikzpicture}}}
+
\raisebox{-10 mm}{\scalebox{0.7}{
\begin{tikzpicture}
  \begin{feynman}[small]
    \vertex (a) {\(1\)};
    \vertex (v1) [right= 1 cm of a];
    \vertex [below right =0.8 cm of v1] (v2);
    \vertex [below right =0.8 cm of v2] (v3);
    \vertex [above right =0.8 cm of v2] (v4);
    \vertex [above right =0.8 cm of v4] (v5);
    \vertex [above right = 0.9 cm of v1] (b) {\(2\)};
    \vertex [right = 0.7 cm of v4] (e) {\(5\)};
    \vertex [below right = 0.7 cm of v4] (f) {\(6\)};
    \vertex [above right = 0.7 cm of v5] (c) {\(3\)};
    \vertex [right = 0.7 cm of v5] (d) {\(4\)};
    \vertex [right = 0.7 cm of v3] (g) {\(7\)};
    \vertex [below right = 0.7 cm of v3] (h) {\(8\)};
    \diagram* {
      (a) -- [solid,line width=0.5mm] (v1) -- [solid,line width=0.5mm] (b),
      (v1) -- [boson,line width=0.5mm] (v2) -- [boson,line width=0.5mm] (v3),
      (v2) -- [boson,line width=0.5mm] (v4) -- [boson,line width=0.5mm] (v5),
      (v5) -- [solid,line width=0.5mm] (c),
      (v5) -- [solid,line width=0.5mm] (d),
      (v4) -- [solid,line width=0.5mm] (e),
      (v4) -- [solid,line width=0.5mm] (f),
      (v3) -- [solid,line width=0.5mm] (g),
      (v3) -- [solid,line width=0.5mm] (h)
    };
  \end{feynman}
\end{tikzpicture}}}
\\
&
+
\raisebox{-10 mm}{\scalebox{0.7}{
\begin{tikzpicture}
  \begin{feynman}[small]
    \vertex (a) {\(1\)};
    \vertex (v1) [right= 1 cm of a];
    \vertex [below right =0.8 cm of v1] (v2);
    \vertex [below right =0.8 cm of v2] (v3);
    \vertex [right =0.8 cm of v3] (v4);
    \vertex [below right =0.8 cm of v3] (v5);
    \vertex [above right = 0.9 cm of v1] (b) {\(2\)};
    \vertex [above right = 0.7 cm of v2] (c) {\(3\)};
    \vertex [right = 0.7 cm of v2] (d) {\(4\)};
    \vertex [above right = 0.7 cm of v4] (e) {\(5\)};
    \vertex [right = 0.7 cm of v4] (f) {\(6\)};
    \vertex [right = 0.7 cm of v5] (g) {\(7\)};
    \vertex [below right = 0.7 cm of v5] (h) {\(8\)};
    \diagram* {
      (a) -- [solid,line width=0.5mm] (v1) -- [solid,line width=0.5mm] (b),
      (v1) -- [boson,line width=0.5mm] (v2) -- [boson,line width=0.5mm] (v3),
      (v3) -- [boson,line width=0.5mm] (v4),
      (v3) -- [boson,line width=0.5mm] (v5),
      (v2) -- [solid,line width=0.5mm] (d),
      (v4) -- [solid,line width=0.5mm] (e),
      (v4) -- [solid,line width=0.5mm] (f),
      (v5) -- [solid,line width=0.5mm] (g),
      (v2) -- [solid,line width=0.5mm] (c),
      (v5) -- [solid,line width=0.5mm] (h)
    };
  \end{feynman}
\end{tikzpicture}}}
+
\raisebox{-10 mm}{\scalebox{0.7}{
\begin{tikzpicture}
  \begin{feynman}[small]
    \vertex (a) {\(1\)};
    \vertex (v1) [right= 1 cm of a];
    \vertex [below right =0.8 cm of v1] (v2);
    \vertex [below right =0.8 cm of v2] (v3);
    \vertex [below right =0.8 cm of v3] (v5);
    \vertex [above right = 0.9 cm of v1] (b) {\(2\)};
    \vertex [above right = 0.7 cm of v2] (c) {\(3\)};
    \vertex [right = 0.7 cm of v2] (d) {\(4\)};
    \vertex [above right = 0.7 cm of v3] (e) {\(5\)};
    \vertex [right = 0.7 cm of v3] (f) {\(6\)};
    \vertex [right = 0.7 cm of v5] (g) {\(7\)};
    \vertex [below right = 0.7 cm of v5] (h) {\(8\)};
    \diagram* {
      (a) -- [solid,line width=0.5mm] (v1) -- [solid,line width=0.5mm] (b),
      (v1) -- [boson,line width=0.5mm] (v2) -- [boson,line width=0.5mm] (v3),
      (v3) -- [boson,line width=0.5mm] (v5),
      (v2) -- [solid,line width=0.5mm] (d),
      (v3) -- [solid,line width=0.5mm] (e),
      (v3) -- [solid,line width=0.5mm] (f),
      (v5) -- [solid,line width=0.5mm] (g),
      (v2) -- [solid,line width=0.5mm] (c),
      (v5) -- [solid,line width=0.5mm] (h)
    };
  \end{feynman}
\end{tikzpicture}}}
+
\raisebox{-10 mm}{\scalebox{0.7}{
\begin{tikzpicture}
  \begin{feynman}[small]
    \vertex (a) {\(1\)};
    \vertex (v1) [right= 1 cm of a];
    \vertex [below right =0.8 cm of v1] (v2);
    \vertex [below right =0.8 cm of v2] (v3);
    \vertex [right =0.8 cm of v3] (v5);
    \vertex [above right = 0.9 cm of v1] (b) {\(2\)};
    \vertex [above right = 0.7 cm of v2] (c) {\(3\)};
    \vertex [right = 0.7 cm of v2] (d) {\(4\)};
    \vertex [above right = 0.7 cm of v3] (e) {\(5\)};
    \vertex [below right = 0.7 cm of v3] (f) {\(8\)};
    \vertex [above right = 0.7 cm of v5] (g) {\(6\)};
    \vertex [below right = 0.7 cm of v5] (h) {\(7\)};
    \diagram* {
      (a) -- [solid,line width=0.5mm] (v1) -- [solid,line width=0.5mm] (b),
      (v1) -- [boson,line width=0.5mm] (v2) -- [boson,line width=0.5mm] (v3),
      (v3) -- [boson,line width=0.5mm] (v5),
      (v2) -- [solid,line width=0.5mm] (d),
      (v3) -- [solid,line width=0.5mm] (e),
      (v3) -- [solid,line width=0.5mm] (f),
      (v5) -- [solid,line width=0.5mm] (g),
      (v2) -- [solid,line width=0.5mm] (c),
      (v5) -- [solid,line width=0.5mm] (h)
    };
  \end{feynman}
\end{tikzpicture}}}
+
\raisebox{-10 mm}{\scalebox{0.7}{
\begin{tikzpicture}
  \begin{feynman}[small]
    \vertex (a) {\(1\)};
    \vertex (v1) [right= 1 cm of a];
    \vertex [below right =0.8 cm of v1] (v2);
    \vertex [below right =0.8 cm of v2] (v3);
    \vertex [right =0.8 cm of v3] (v5);
    \vertex [above right = 0.9 cm of v1] (b) {\(2\)};
    \vertex [above right = 0.7 cm of v2] (c) {\(3\)};
    \vertex [right = 0.7 cm of v2] (d) {\(4\)};
    \vertex [below = 0.7 cm of v3] (e) {\(8\)};
    \vertex [below right = 0.7 cm of v3] (f) {\(7\)};
    \vertex [above right = 0.7 cm of v5] (g) {\(5\)};
    \vertex [below right = 0.7 cm of v5] (h) {\(6\)};
    \diagram* {
      (a) -- [solid,line width=0.5mm] (v1) -- [solid,line width=0.5mm] (b),
      (v1) -- [boson,line width=0.5mm] (v2) -- [boson,line width=0.5mm] (v3),
      (v3) -- [boson,line width=0.5mm] (v5),
      (v2) -- [solid,line width=0.5mm] (d),
      (v3) -- [solid,line width=0.5mm] (e),
      (v3) -- [solid,line width=0.5mm] (f),
      (v5) -- [solid,line width=0.5mm] (g),
      (v2) -- [solid,line width=0.5mm] (c),
      (v5) -- [solid,line width=0.5mm] (h)
    };
  \end{feynman}
\end{tikzpicture}}}
\\
&
+
\raisebox{-10 mm}{\scalebox{0.7}{
\begin{tikzpicture}
  \begin{feynman}[small]
    \vertex (a) {\(1\)};
    \vertex (v1) [right= 1 cm of a];
    \vertex [below right =0.8 cm of v1] (v2);
    \vertex [right =0.8 cm of v2] (v3);
    \vertex [above right =0.8 cm of v3] (v4);
    \vertex [below right =0.8 cm of v3] (v5);
    \vertex [above right = 0.9 cm of v1] (b) {\(2\)};
    \vertex [above right = 0.7 cm of v2] (c) {\(3\)};
    \vertex [below right = 0.7 cm of v2] (h) {\(8\)};
    \vertex [above right = 0.7 cm of v4] (d) {\(4\)};
    \vertex [right = 0.7 cm of v4] (e) {\(5\)};
    \vertex [right = 0.7 cm of v5] (f) {\(6\)};
    \vertex [below right = 0.7 cm of v5] (g) {\(7\)};
    \diagram* {
      (a) -- [solid,line width=0.5mm] (v1) -- [solid,line width=0.5mm] (b),
      (v1) -- [boson,line width=0.5mm] (v2) -- [boson,line width=0.5mm] (v3),
      (v3) -- [boson,line width=0.5mm] (v4),
      (v3) -- [boson,line width=0.5mm] (v5),
      (v4) -- [solid,line width=0.5mm] (d),
      (v4) -- [solid,line width=0.5mm] (e),
      (v5) -- [solid,line width=0.5mm] (f),
      (v5) -- [solid,line width=0.5mm] (g),
      (v2) -- [solid,line width=0.5mm] (c),
      (v2) -- [solid,line width=0.5mm] (h)
    };
  \end{feynman}
\end{tikzpicture}}}
+
\raisebox{-10 mm}{\scalebox{0.7}{
\begin{tikzpicture}
  \begin{feynman}[small]
    \vertex (a) {\(1\)};
    \vertex (v1) [right= 1 cm of a];
    \vertex [below right =0.8 cm of v1] (v2);
    \vertex [right =0.8 cm of v2] (v3);
    \vertex [below right =0.8 cm of v3] (v5);
    \vertex [above right = 0.9 cm of v1] (b) {\(2\)};
    \vertex [above right = 0.7 cm of v2] (c) {\(3\)};
    \vertex [below right = 0.7 cm of v2] (h) {\(8\)};
    \vertex [above right = 0.7 cm of v3] (d) {\(4\)};
    \vertex [right = 0.7 cm of v3] (e) {\(5\)};
    \vertex [above right = 0.7 cm of v5] (f) {\(6\)};
    \vertex [below right = 0.7 cm of v5] (g) {\(7\)};
    \diagram* {
      (a) -- [solid,line width=0.5mm] (v1) -- [solid,line width=0.5mm] (b),
      (v1) -- [boson,line width=0.5mm] (v2) -- [boson,line width=0.5mm] (v3),
      (v3) -- [boson,line width=0.5mm] (v5),
      (v3) -- [solid,line width=0.5mm] (d),
      (v3) -- [solid,line width=0.5mm] (e),
      (v5) -- [solid,line width=0.5mm] (f),
      (v5) -- [solid,line width=0.5mm] (g),
      (v2) -- [solid,line width=0.5mm] (c),
      (v2) -- [solid,line width=0.5mm] (h)
    };
  \end{feynman}
\end{tikzpicture}}}
+
\raisebox{-10 mm}{\scalebox{0.7}{
\begin{tikzpicture}
  \begin{feynman}[small]
    \vertex (a) {\(1\)};
    \vertex (v1) [right= 1 cm of a];
    \vertex [below right =0.8 cm of v1] (v2);
    \vertex [right =0.8 cm of v2] (v3);
    \vertex [right =0.8 cm of v3] (v5);
    \vertex [above right = 0.9 cm of v1] (b) {\(2\)};
    \vertex [above right = 0.7 cm of v2] (c) {\(3\)};
    \vertex [below right = 0.7 cm of v2] (h) {\(8\)};
    \vertex [above right = 0.7 cm of v3] (d) {\(4\)};
    \vertex [below right = 0.7 cm of v3] (e) {\(7\)};
    \vertex [above right = 0.7 cm of v5] (f) {\(5\)};
    \vertex [below right = 0.7 cm of v5] (g) {\(6\)};
    \diagram* {
      (a) -- [solid,line width=0.5mm] (v1) -- [solid,line width=0.5mm] (b),
      (v1) -- [boson,line width=0.5mm] (v2) -- [boson,line width=0.5mm] (v3),
      (v3) -- [boson,line width=0.5mm] (v5),
      (v3) -- [solid,line width=0.5mm] (d),
      (v3) -- [solid,line width=0.5mm] (e),
      (v5) -- [solid,line width=0.5mm] (f),
      (v5) -- [solid,line width=0.5mm] (g),
      (v2) -- [solid,line width=0.5mm] (c),
      (v2) -- [solid,line width=0.5mm] (h)
    };
  \end{feynman}
\end{tikzpicture}}}
+
\raisebox{-10 mm}{\scalebox{0.7}{
\begin{tikzpicture}
  \begin{feynman}[small]
    \vertex (a) {\(1\)};
    \vertex (v1) [right= 1 cm of a];
    \vertex [below right =0.8 cm of v1] (v2);
    \vertex [right =0.8 cm of v2] (v3);
    \vertex [above right =0.8 cm of v3] (v5);
    \vertex [above right = 0.9 cm of v1] (b) {\(2\)};
    \vertex [above right = 0.7 cm of v2] (c) {\(3\)};
    \vertex [below right = 0.7 cm of v2] (h) {\(8\)};
    \vertex [right = 0.7 cm of v3] (d) {\(6\)};
    \vertex [below right = 0.7 cm of v3] (e) {\(7\)};
    \vertex [above right = 0.7 cm of v5] (f) {\(4\)};
    \vertex [below right = 0.7 cm of v5] (g) {\(5\)};
    \diagram* {
      (a) -- [solid,line width=0.5mm] (v1) -- [solid,line width=0.5mm] (b),
      (v1) -- [boson,line width=0.5mm] (v2) -- [boson,line width=0.5mm] (v3),
      (v3) -- [boson,line width=0.5mm] (v5),
      (v3) -- [solid,line width=0.5mm] (d),
      (v3) -- [solid,line width=0.5mm] (e),
      (v5) -- [solid,line width=0.5mm] (f),
      (v5) -- [solid,line width=0.5mm] (g),
      (v2) -- [solid,line width=0.5mm] (c),
      (v2) -- [solid,line width=0.5mm] (h)
    };
  \end{feynman}
\end{tikzpicture}}}
\\
&
+
\raisebox{-10 mm}{\scalebox{0.7}{
\begin{tikzpicture}
  \begin{feynman}[small]
    \vertex (a) {\(1\)};
    \vertex (v1) [right= 1 cm of a];
    \vertex [below right =0.8 cm of v1] (v2);
    \vertex [above right =0.8 cm of v2] (v3);
    \vertex [above right =0.8 cm of v3] (v4);
    \vertex [right =0.8 cm of v3] (v5);
    \vertex [above right = 0.9 cm of v1] (b) {\(2\)};
    \vertex [right = 0.7 cm of v2] (g) {\(7\)};
    \vertex [below right = 0.7 cm of v2] (h) {\(8\)};
    \vertex [above right = 0.7 cm of v4] (c) {\(3\)};
    \vertex [right = 0.7 cm of v4] (d) {\(4\)};
    \vertex [right = 0.7 cm of v5] (e) {\(5\)};
    \vertex [below right = 0.7 cm of v5] (f) {\(6\)};
    \diagram* {
      (a) -- [solid,line width=0.5mm] (v1) -- [solid,line width=0.5mm] (b),
      (v1) -- [boson,line width=0.5mm] (v2) -- [boson,line width=0.5mm] (v3),
      (v3) -- [boson,line width=0.5mm] (v4),
      (v3) -- [boson,line width=0.5mm] (v5),
      (v4) -- [solid,line width=0.5mm] (d),
      (v5) -- [solid,line width=0.5mm] (e),
      (v5) -- [solid,line width=0.5mm] (f),
      (v2) -- [solid,line width=0.5mm] (g),
      (v4) -- [solid,line width=0.5mm] (c),
      (v2) -- [solid,line width=0.5mm] (h)
    };
  \end{feynman}
\end{tikzpicture}}}
+
\raisebox{-10 mm}{\scalebox{0.7}{
\begin{tikzpicture}
  \begin{feynman}[small]
    \vertex (a) {\(1\)};
    \vertex (v1) [right= 1 cm of a];
    \vertex [below right =0.8 cm of v1] (v2);
    \vertex [above right =0.8 cm of v2] (v3);
    \vertex [right =0.8 cm of v3] (v5);
    \vertex [above right = 0.9 cm of v1] (b) {\(2\)};
    \vertex [right = 0.7 cm of v2] (g) {\(7\)};
    \vertex [below right = 0.7 cm of v2] (h) {\(8\)};
    \vertex [above = 0.7 cm of v3] (c) {\(3\)};
    \vertex [above right = 0.7 cm of v3] (d) {\(4\)};
    \vertex [right = 0.7 cm of v5] (e) {\(5\)};
    \vertex [below right = 0.7 cm of v5] (f) {\(6\)};
    \diagram* {
      (a) -- [solid,line width=0.5mm] (v1) -- [solid,line width=0.5mm] (b),
      (v1) -- [boson,line width=0.5mm] (v2) -- [boson,line width=0.5mm] (v3),
      (v3) -- [boson,line width=0.5mm] (v5),
      (v3) -- [solid,line width=0.5mm] (d),
      (v5) -- [solid,line width=0.5mm] (e),
      (v5) -- [solid,line width=0.5mm] (f),
      (v2) -- [solid,line width=0.5mm] (g),
      (v3) -- [solid,line width=0.5mm] (c),
      (v2) -- [solid,line width=0.5mm] (h)
    };
  \end{feynman}
\end{tikzpicture}}}
+
\raisebox{-10 mm}{\scalebox{0.7}{
\begin{tikzpicture}
  \begin{feynman}[small]
    \vertex (a) {\(1\)};
    \vertex (v1) [right= 1 cm of a];
    \vertex [below right =0.8 cm of v1] (v2);
    \vertex [above right =0.8 cm of v2] (v3);
    \vertex [above right =0.8 cm of v3] (v5);
    \vertex [above right = 0.9 cm of v1] (b) {\(2\)};
    \vertex [right = 0.7 cm of v2] (g) {\(7\)};
    \vertex [below right = 0.7 cm of v2] (h) {\(8\)};
    \vertex [above = 0.7 cm of v3] (c) {\(3\)};
    \vertex [right = 0.7 cm of v3] (d) {\(6\)};
    \vertex [above right = 0.7 cm of v5] (e) {\(4\)};
    \vertex [right = 0.7 cm of v5] (f) {\(5\)};
    \diagram* {
      (a) -- [solid,line width=0.5mm] (v1) -- [solid,line width=0.5mm] (b),
      (v1) -- [boson,line width=0.5mm] (v2) -- [boson,line width=0.5mm] (v3),
      (v3) -- [boson,line width=0.5mm] (v5),
      (v3) -- [solid,line width=0.5mm] (d),
      (v5) -- [solid,line width=0.5mm] (e),
      (v5) -- [solid,line width=0.5mm] (f),
      (v2) -- [solid,line width=0.5mm] (g),
      (v3) -- [solid,line width=0.5mm] (c),
      (v2) -- [solid,line width=0.5mm] (h)
    };
  \end{feynman}
\end{tikzpicture}}}
+
\raisebox{-10 mm}{\scalebox{0.7}{
\begin{tikzpicture}
  \begin{feynman}[small]
    \vertex (a) {\(1\)};
    \vertex (v1) [right= 1 cm of a];
    \vertex [below right =0.8 cm of v1] (v2);
    \vertex [above right =0.8 cm of v2] (v3);
    \vertex [above =0.8 cm of v3] (v5);
    \vertex [above right = 0.9 cm of v1] (b) {\(2\)};
    \vertex [right = 0.7 cm of v2] (g) {\(7\)};
    \vertex [below right = 0.7 cm of v2] (h) {\(8\)};
    \vertex [above right = 0.7 cm of v3] (c) {\(5\)};
    \vertex [right = 0.7 cm of v3] (d) {\(6\)};
    \vertex [above right = 0.7 cm of v5] (e) {\(3\)};
    \vertex [right = 0.7 cm of v5] (f) {\(4\)};
    \diagram* {
      (a) -- [solid,line width=0.5mm] (v1) -- [solid,line width=0.5mm] (b),
      (v1) -- [boson,line width=0.5mm] (v2) -- [boson,line width=0.5mm] (v3),
      (v3) -- [boson,line width=0.5mm] (v5),
      (v3) -- [solid,line width=0.5mm] (d),
      (v5) -- [solid,line width=0.5mm] (e),
      (v5) -- [solid,line width=0.5mm] (f),
      (v2) -- [solid,line width=0.5mm] (g),
      (v3) -- [solid,line width=0.5mm] (c),
      (v2) -- [solid,line width=0.5mm] (h)
    };
  \end{feynman}
\end{tikzpicture}}}
\\
&
+
\raisebox{-10 mm}{\scalebox{0.7}{
\begin{tikzpicture}
  \begin{feynman}[small]
    \vertex (a) {\(1\)};
    \vertex (v1) [right= 1 cm of a];
    \vertex [right =0.8 cm of v1] (v2);
    \vertex [below right =0.8 cm of v1] (v3);
    \vertex [right =0.8 cm of v3] (v4);
    \vertex [below =0.8 cm of v3] (v5);
    \vertex [above right = 0.9 cm of v1] (b) {\(2\)};
    \vertex [above right = 0.7 cm of v2] (c) {\(3\)};
    \vertex [right = 0.5 cm of v2] (d) {\(4\)};
    \vertex [right = 0.7 cm of v4] (e) {\(5\)};
    \vertex [below right = 0.7 cm of v4] (f) {\(6\)};
    \vertex [right = 0.7 cm of v5] (g) {\(7\)};
    \vertex [below right = 0.7 cm of v5] (h) {\(8\)};
    \diagram* {
      (a) -- [solid,line width=0.5mm] (v1) -- [solid,line width=0.5mm] (b),
      (v1) -- [boson,line width=0.5mm] (v2),
      (v1) -- [boson,line width=0.5mm] (v3)  -- [boson,line width=0.5mm] (v4),
      (v3) -- [boson,line width=0.5mm] (v5),
      (v2) -- [solid,line width=0.5mm] (c),
      (v2) -- [solid,line width=0.5mm] (d),
      (v4) -- [solid,line width=0.5mm] (e),
      (v4) -- [solid,line width=0.5mm] (f),
      (v5) -- [solid,line width=0.5mm] (g),
      (v5) -- [solid,line width=0.5mm] (h)
    };
  \end{feynman}
\end{tikzpicture}}}
+
\raisebox{-10 mm}{\scalebox{0.7}{
\begin{tikzpicture}
  \begin{feynman}[small]
    \vertex (a) {\(1\)};
    \vertex (v1) [right= 1 cm of a];
    \vertex [right =0.8 cm of v1] (v2);
    \vertex [below=0.8 cm of v1] (v3);
    \vertex [below=0.8 cm of v3] (v4);
    \vertex [above right = 0.9 cm of v1] (b) {\(2\)};
    \vertex [above right = 0.7 cm of v2] (c) {\(3\)};
    \vertex [right = 0.7 cm of v2] (d) {\(4\)};
    \vertex [below right = 0.7 cm of v3] (e) {\(6\)};
    \vertex [right = 0.7 cm of v3] (h) {\(5\)};
    \vertex [right = 0.7 cm of v4] (f) {\(7\)};
    \vertex [below right = 0.7 cm of v4] (g) {\(8\)};
    \diagram* {
      (a) -- [solid,line width=0.5mm] (v1) -- [solid,line width=0.5mm] (b),
      (v1) -- [boson,line width=0.5mm] (v2),
      (v1) -- [boson,line width=0.5mm] (v3)  -- [boson,line width=0.5mm] (v4),
      (v2) -- [solid,line width=0.5mm] (c),
      (v2) -- [solid,line width=0.5mm] (d),
      (v3) -- [solid,line width=0.5mm] (e),
      (v4) -- [solid,line width=0.5mm] (f),
      (v4) -- [solid,line width=0.5mm] (g),
      (v3) -- [solid,line width=0.5mm] (h)
    };
  \end{feynman}
\end{tikzpicture}}}
+
\raisebox{-10 mm}{\scalebox{0.7}{
\begin{tikzpicture}
  \begin{feynman}[small]
    \vertex (a) {\(1\)};
    \vertex (v1) [right= 1 cm of a];
    \vertex [right =0.8 cm of v1] (v2);
    \vertex [below=0.8 cm of v1] (v3);
    \vertex [below right =0.8 cm of v3] (v4);
    \vertex [above right = 0.9 cm of v1] (b) {\(2\)};
    \vertex [above right = 0.7 cm of v2] (c) {\(3\)};
    \vertex [right = 0.7 cm of v2] (d) {\(4\)};
    \vertex [right = 0.7 cm of v3] (e) {\(5\)};
    \vertex [below = 0.7 cm of v3] (h) {\(8\)};
    \vertex [right = 0.7 cm of v4] (f) {\(6\)};
    \vertex [below right = 0.7 cm of v4] (g) {\(7\)};
    \diagram* {
      (a) -- [solid,line width=0.5mm] (v1) -- [solid,line width=0.5mm] (b),
      (v1) -- [boson,line width=0.5mm] (v2),
      (v1) -- [boson,line width=0.5mm] (v3)  -- [boson,line width=0.5mm] (v4),
      (v2) -- [solid,line width=0.5mm] (c),
      (v2) -- [solid,line width=0.5mm] (d),
      (v3) -- [solid,line width=0.5mm] (e),
      (v4) -- [solid,line width=0.5mm] (f),
      (v4) -- [solid,line width=0.5mm] (g),
      (v3) -- [solid,line width=0.5mm] (h)
    };
  \end{feynman}
\end{tikzpicture}}}
+
\raisebox{-10 mm}{\scalebox{0.7}{
\begin{tikzpicture}
  \begin{feynman}[small]
    \vertex (a) {\(1\)};
    \vertex (v1) [right= 1 cm of a];
    \vertex [right =0.8 cm of v1] (v2);
    \vertex [below=0.8 cm of v1] (v3);
    \vertex [right =0.8 cm of v3] (v4);
    \vertex [above right = 0.9 cm of v1] (b) {\(2\)};
    \vertex [above right = 0.7 cm of v2] (c) {\(3\)};
    \vertex [right = 0.7 cm of v2] (d) {\(4\)};
    \vertex [below right = 0.7 cm of v3] (e) {\(7\)};
    \vertex [below = 0.7 cm of v3] (h) {\(8\)};
    \vertex [right = 0.7 cm of v4] (f) {\(5\)};
    \vertex [below right = 0.7 cm of v4] (g) {\(6\)};
    \diagram* {
      (a) -- [solid,line width=0.5mm] (v1) -- [solid,line width=0.5mm] (b),
      (v1) -- [boson,line width=0.5mm] (v2),
      (v1) -- [boson,line width=0.5mm] (v3)  -- [boson,line width=0.5mm] (v4),
      (v2) -- [solid,line width=0.5mm] (c),
      (v2) -- [solid,line width=0.5mm] (d),
      (v3) -- [solid,line width=0.5mm] (e),
      (v4) -- [solid,line width=0.5mm] (f),
      (v4) -- [solid,line width=0.5mm] (g),
      (v3) -- [solid,line width=0.5mm] (h)
    };
  \end{feynman}
\end{tikzpicture}}}
\\
&
+
\raisebox{-10 mm}{\scalebox{0.7}{
\begin{tikzpicture}
  \begin{feynman}[small]
    \vertex (a) {\(1\)};
    \vertex (v1) [right= 1 cm of a];
    \vertex [below right =0.8 cm of v1] (v2);
    \vertex [right =0.8 cm of v1] (v3);
    \vertex [above right =0.8 cm of v3] (v4);
    \vertex [right =0.8 cm of v3] (v5);
    \vertex [above right = 0.9 cm of v1] (b) {\(2\)};
    \vertex [right = 0.7 cm of v2] (g) {\(7\)};
    \vertex [below right = 0.7 cm of v2] (h) {\(8\)};
    \vertex [above right = 0.7 cm of v4] (c) {\(3\)};
    \vertex [right = 0.7 cm of v4] (d) {\(4\)};
    \vertex [right = 0.7 cm of v5] (e) {\(5\)};
    \vertex [below right = 0.7 cm of v5] (f) {\(6\)};
    \diagram* {
      (a) -- [solid,line width=0.5mm] (v1) -- [solid,line width=0.5mm] (b),
      (v1) -- [boson,line width=0.5mm] (v2),
      (v1) -- [boson,line width=0.5mm] (v3)  -- [boson,line width=0.5mm] (v4),
      (v3) -- [boson,line width=0.5mm] (v5),
      (v4) -- [solid,line width=0.5mm] (c),
      (v4) -- [solid,line width=0.5mm] (d),
      (v5) -- [solid,line width=0.5mm] (e),
      (v5) -- [solid,line width=0.5mm] (f),
      (v2) -- [solid,line width=0.5mm] (g),
      (v2) -- [solid,line width=0.5mm] (h)
    };
  \end{feynman}
\end{tikzpicture}}}
+
\raisebox{-10 mm}{\scalebox{0.7}{
\begin{tikzpicture}
  \begin{feynman}[small]
    \vertex (a) {\(1\)};
    \vertex (v1) [right= 1 cm of a];
    \vertex [below =0.8 cm of v1] (v2);
    \vertex [right =0.8 cm of v1] (v3);
    \vertex [below right =0.8 cm of v3] (v4);
    \vertex [above right = 0.9 cm of v1] (b) {\(2\)};
    \vertex [right = 0.7 cm of v2] (g) {\(7\)};
    \vertex [below right = 0.7 cm of v2] (h) {\(8\)};
    \vertex [above right = 0.7 cm of v3] (c) {\(3\)};
    \vertex [right = 0.7 cm of v3] (f) {\(4\)};
    \vertex [above right = 0.7 cm of v4] (d) {\(5\)};
    \vertex [below right = 0.7 cm of v4] (e) {\(6\)};
    \diagram* {
      (a) -- [solid,line width=0.5mm] (v1) -- [solid,line width=0.5mm] (b),
      (v1) -- [boson,line width=0.5mm] (v2),
      (v1) -- [boson,line width=0.5mm] (v3)  -- [boson,line width=0.5mm] (v4),
      (v3) -- [solid,line width=0.5mm] (c),
      (v4) -- [solid,line width=0.5mm] (d),
      (v4) -- [solid,line width=0.5mm] (e),
      (v3) -- [solid,line width=0.5mm] (f),
      (v2) -- [solid,line width=0.5mm] (g),
      (v2) -- [solid,line width=0.5mm] (h)
    };
  \end{feynman}
\end{tikzpicture}}}
+
\raisebox{-10 mm}{\scalebox{0.7}{
\begin{tikzpicture}
  \begin{feynman}[small]
    \vertex (a) {\(1\)};
    \vertex (v1) [right= 1 cm of a];
    \vertex [below =0.8 cm of v1] (v2);
    \vertex [right =0.8 cm of v1] (v3);
    \vertex [right =0.8 cm of v3] (v4);
    \vertex [above right = 0.9 cm of v1] (b) {\(2\)};
    \vertex [right = 0.7 cm of v2] (g) {\(7\)};
    \vertex [below right = 0.7 cm of v2] (h) {\(8\)};
    \vertex [above right = 0.7 cm of v3] (c) {\(3\)};
    \vertex [below right = 0.7 cm of v3] (f) {\(6\)};
    \vertex [above right = 0.7 cm of v4] (d) {\(4\)};
    \vertex [below right = 0.7 cm of v4] (e) {\(5\)};
    \diagram* {
      (a) -- [solid,line width=0.5mm] (v1) -- [solid,line width=0.5mm] (b),
      (v1) -- [boson,line width=0.5mm] (v2),
      (v1) -- [boson,line width=0.5mm] (v3)  -- [boson,line width=0.5mm] (v4),
      (v3) -- [solid,line width=0.5mm] (c),
      (v4) -- [solid,line width=0.5mm] (d),
      (v4) -- [solid,line width=0.5mm] (e),
      (v3) -- [solid,line width=0.5mm] (f),
      (v2) -- [solid,line width=0.5mm] (g),
      (v2) -- [solid,line width=0.5mm] (h)
    };
  \end{feynman}
\end{tikzpicture}}}
+
\raisebox{-10 mm}{\scalebox{0.7}{
\begin{tikzpicture}
  \begin{feynman}[small]
    \vertex (a) {\(1\)};
    \vertex (v1) [right= 1 cm of a];
    \vertex [below =0.8 cm of v1] (v2);
    \vertex [right =0.8 cm of v1] (v3);
    \vertex [above right =0.8 cm of v3] (v4);
    \vertex [above right = 0.9 cm of v1] (b) {\(2\)};
    \vertex [right = 0.7 cm of v2] (g) {\(7\)};
    \vertex [below right = 0.7 cm of v2] (h) {\(8\)};
    \vertex [right = 0.7 cm of v3] (c) {\(5\)};
    \vertex [below right = 0.7 cm of v3] (f) {\(6\)};
    \vertex [above right = 0.7 cm of v4] (d) {\(3\)};
    \vertex [right = 0.7 cm of v4] (e) {\(4\)};
    \diagram* {
      (a) -- [solid,line width=0.5mm] (v1) -- [solid,line width=0.5mm] (b),
      (v1) -- [boson,line width=0.5mm] (v2),
      (v1) -- [boson,line width=0.5mm] (v3)  -- [boson,line width=0.5mm] (v4),
      (v3) -- [solid,line width=0.5mm] (c),
      (v4) -- [solid,line width=0.5mm] (d),
      (v4) -- [solid,line width=0.5mm] (e),
      (v3) -- [solid,line width=0.5mm] (f),
      (v2) -- [solid,line width=0.5mm] (g),
      (v2) -- [solid,line width=0.5mm] (h)
    };
  \end{feynman}
\end{tikzpicture}}}
\\
&
+
\raisebox{-10 mm}{\scalebox{0.7}{
\begin{tikzpicture}
  \begin{feynman}[small]
    \vertex (a) {\(1\)};
    \vertex (v1) [right= 1 cm of a];
    \vertex [below right =0.8 cm of v1] (v2);
    \vertex [above right =0.8 cm of v1] (v3);
    \vertex [above right =0.8 cm of v3] (v4);
    \vertex [right =0.8 cm of v3] (v5);
    \vertex [right = 1.2 cm of v1] (f) {\(6\)};
    \vertex [right = 0.7 cm of v2] (g) {\(7\)};
    \vertex [below right = 0.7 cm of v2] (h) {\(8\)};
    \vertex [above right = 0.7 cm of v4] (b) {\(2\)};
    \vertex [right = 0.7 cm of v4] (c) {\(3\)};
    \vertex [right = 0.7 cm of v5] (d) {\(4\)};
    \vertex [below right = 0.7 cm of v5] (e) {\(5\)};
    \diagram* {
      (a) -- [solid,line width=0.5mm] (v1) -- [solid,line width=0.5mm] (f),
      (v1) -- [boson,line width=0.5mm] (v2),
      (v1) -- [boson,line width=0.5mm] (v3)  -- [boson,line width=0.5mm] (v4),
      (v3) -- [boson,line width=0.5mm] (v5),
      (v4) -- [solid,line width=0.5mm] (b),
      (v4) -- [solid,line width=0.5mm] (c),
      (v5) -- [solid,line width=0.5mm] (d),
      (v5) -- [solid,line width=0.5mm] (e),
      (v2) -- [solid,line width=0.5mm] (g),
      (v2) -- [solid,line width=0.5mm] (h)
    };
  \end{feynman}
\end{tikzpicture}}}
+
\raisebox{-10 mm}{\scalebox{0.7}{
\begin{tikzpicture}
  \begin{feynman}[small]
    \vertex (a) {\(1\)};
    \vertex (v1) [right= 1 cm of a];
    \vertex [above right =0.8 cm of v1] (v3);
    \vertex [below right =0.8 cm of v1] (v2);
    \vertex [right =0.8 cm of v3] (v4);
    \vertex [right = 1.2 cm of v1] (f) {\(6\)};
    \vertex [right = 0.7 cm of v2] (g) {\(7\)};
    \vertex [below right = 0.7 cm of v2] (h) {\(8\)};
    \vertex [above = 0.7 cm of v3] (b) {\(2\)};
    \vertex [above right = 0.7 cm of v3] (e) {\(3\)};
    \vertex [above right = 0.7 cm of v4] (c) {\(4\)};
    \vertex [right = 0.7 cm of v4] (d) {\(5\)};
    \diagram* {
      (a) -- [solid,line width=0.5mm] (v1) -- [solid,line width=0.5mm] (f),
      (v1) -- [boson,line width=0.5mm] (v2),
      (v1) -- [boson,line width=0.5mm] (v3)  -- [boson,line width=0.5mm] (v4),
      (v4) -- [solid,line width=0.5mm] (c),
      (v4) -- [solid,line width=0.5mm] (d),
      (v3) -- [solid,line width=0.5mm] (e),
      (v3) -- [solid,line width=0.5mm] (b),
      (v2) -- [solid,line width=0.5mm] (g),
      (v2) -- [solid,line width=0.5mm] (h)
    };
  \end{feynman}
\end{tikzpicture}}}
+
\raisebox{-10 mm}{\scalebox{0.7}{
\begin{tikzpicture}
  \begin{feynman}[small]
    \vertex (a) {\(1\)};
    \vertex (v1) [right= 1 cm of a];
    \vertex [above right =0.8 cm of v1] (v3);
    \vertex [below right =0.8 cm of v1] (v2);
    \vertex [above right =0.8 cm of v3] (v4);
    \vertex [right = 1.2 cm of v1] (f) {\(6\)};
    \vertex [right = 0.7 cm of v2] (g) {\(7\)};
    \vertex [below right = 0.7 cm of v2] (h) {\(8\)};
    \vertex [above = 0.7 cm of v3] (b) {\(2\)};
    \vertex [right = 0.7 cm of v3] (e) {\(5\)};
    \vertex [above right = 0.7 cm of v4] (c) {\(3\)};
    \vertex [right = 0.7 cm of v4] (d) {\(4\)};
    \diagram* {
      (a) -- [solid,line width=0.5mm] (v1) -- [solid,line width=0.5mm] (f),
      (v1) -- [boson,line width=0.5mm] (v2),
      (v1) -- [boson,line width=0.5mm] (v3)  -- [boson,line width=0.5mm] (v4),
      (v4) -- [solid,line width=0.5mm] (c),
      (v4) -- [solid,line width=0.5mm] (d),
      (v3) -- [solid,line width=0.5mm] (e),
      (v3) -- [solid,line width=0.5mm] (b),
      (v2) -- [solid,line width=0.5mm] (g),
      (v2) -- [solid,line width=0.5mm] (h)
    };
  \end{feynman}
\end{tikzpicture}}}
+
\raisebox{-10 mm}{\scalebox{0.7}{
\begin{tikzpicture}
  \begin{feynman}[small]
    \vertex (a) {\(1\)};
    \vertex (v1) [right= 1 cm of a];
    \vertex [above right =0.8 cm of v1] (v3);
    \vertex [below right =0.8 cm of v1] (v2);
    \vertex [above =0.8 cm of v3] (v4);
    \vertex [right = 1.2 cm of v1] (f) {\(6\)};
    \vertex [right = 0.7 cm of v2] (g) {\(7\)};
    \vertex [below right = 0.7 cm of v2] (h) {\(8\)};
    \vertex [above right = 0.7 cm of v3] (b) {\(4\)};
    \vertex [right = 0.7 cm of v3] (e) {\(5\)};
    \vertex [above right = 0.7 cm of v4] (c) {\(2\)};
    \vertex [right = 0.7 cm of v4] (d) {\(3\)};
    \diagram* {
      (a) -- [solid,line width=0.5mm] (v1) -- [solid,line width=0.5mm] (f),
      (v1) -- [boson,line width=0.5mm] (v2),
      (v1) -- [boson,line width=0.5mm] (v3)  -- [boson,line width=0.5mm] (v4),
      (v4) -- [solid,line width=0.5mm] (c),
      (v4) -- [solid,line width=0.5mm] (d),
      (v3) -- [solid,line width=0.5mm] (e),
      (v3) -- [solid,line width=0.5mm] (b),
      (v2) -- [solid,line width=0.5mm] (g),
      (v2) -- [solid,line width=0.5mm] (h)
    };
  \end{feynman}
\end{tikzpicture}}}
\\
&
+
\raisebox{-10 mm}{\scalebox{0.7}{
\begin{tikzpicture}
  \begin{feynman}[small]
    \vertex (a) {\(1\)};
    \vertex (v1) [right= 1 cm of a];
    \vertex [below right =0.8 cm of v1] (v2);
    \vertex [above right =0.8 cm of v2] (v3);
    \vertex [right =0.8 cm of v2] (v4);
    \vertex [below right =0.8 cm of v2] (v5);
    \vertex [above right = 0.9 cm of v1] (b) {\(2\)};
    \vertex [above right = 0.7 cm of v3] (c) {\(3\)};
    \vertex [right = 0.7 cm of v3] (d) {\(4\)};
    \vertex [above right = 0.2cm and 0.8cm of v4] (e) {\(5\)};
    \vertex [below right = 0.2cm and 0.8cm of v4] (f) {\(6\)};
    \vertex [right = 0.7 cm of v5] (g) {\(7\)};
    \vertex [below right = 0.7 cm of v5] (h) {\(8\)};
    \diagram* {
      (a) -- [solid,line width=0.5mm] (v1) -- [solid,line width=0.5mm] (b),
      (v1) -- [boson,line width=0.5mm] (v2) -- [boson,line width=0.5mm] (v3),
      (v2) -- [boson,line width=0.5mm] (v4),
      (v2) -- [boson,line width=0.5mm] (v5),
      (v4) -- [solid,line width=0.5mm] (e),
      (v4) -- [solid,line width=0.5mm] (f),
      (v5) -- [solid,line width=0.5mm] (g),
      (v5) -- [solid,line width=0.5mm] (h),
      (v3) -- [solid,line width=0.5mm] (c),
      (v3) -- [solid,line width=0.5mm] (d)
    };
  \end{feynman}
\end{tikzpicture}}}
\,\,\, + \,\,\, \normalsize{\text{mirror.}}
\end{align*}
\rev{The associated vertex rules corresponding to these graphs are, respectively,
\begin{equation}\hspace{-6.6cm}
 V_4(1,2,3,4) = - (1 + 2 p_2 \cdot (p_3-p_4) / s_{34} ) \delta_{12} \delta_{34} + \delta_{13} \delta_{24} +\text{mirror}, 
\end{equation}
\begin{align}\hspace{-4.5cm}
\notag
 V_6(1, \ldots, 6)=& - \frac{\delta_{12}\delta_{34}\delta_{56}}{s_{34}s_{56}} 
 \left(
    \begin{aligned}
     &-(p_3 - p_4)\cdot(p_5 - p_6)+\tfrac{2}{s_{3456}}\times\\
     &\quad
     \left(
     \begin{aligned}
     &+ p_2 \cdot p_3  (p_3 + 3 p_4) \cdot (p_5 - p_6) \\
     &- p_2 \cdot p_4  (3 p_3 + p_4) \cdot (p_5 - p_6) \\
     &- p_2 \cdot p_5  (p_3 - p_4) \cdot (p_5 + 3 p_6) \\
     &+ p_2 \cdot p_6  (p_3 - p_4) \cdot (3 p_5 + p_6)
     \end{aligned}
     \right)
    \end{aligned}
 \right)\\
 & +\frac{2\delta_{12}}{s_{3456}}
 \left(
 \begin{aligned}
     &+2\delta_{36}\delta_{45} p_2\cdot (p_4-p_5)/s_{45} \\
     &- \delta_{34}\delta_{56}p_2\cdot (p_3-p_4)/s_{34} \\
     &- \delta_{34}\delta_{56} p_2 \cdot (p_5-p_6)/s_{56}
 \end{aligned} 
 \right)- \frac{\delta_{14}\delta_{23}\delta_{56}}{s_{23}s_{56}}(p_2 - p_3)\cdot(p_5 - p_6) + \text{mirror},
\end{align}}
\begin{align}
        \notag V_8(1,\dots,8)=&+ \delta_{14}\times\frac{2\delta_{23}\delta_{56}\delta_{78}}{s_{23}s_{56}s_{78}s_{5678}}\left(\mathcal{U}(2,5,6,7,8)-\mathcal{U}(3,5,6,7,8)\right)\\\notag
        &+
        \delta_{14}\times
        \left(
            \frac{4\delta_{23}\delta_{58}\delta_{67}}{s_{23}s_{67}s_{5678}}(p_2-p_3)\cdot(p_6-p_7)+(678 \to 786) \,\,\mathrm{\&}\,\, (567 \to 756)
        \right)
        \\\notag
        &-\delta_{12}\times\frac{2\delta_{34}\delta_{56}\delta_{78}}{s_{34}s_{56}s_{78}s_{345678}}
            \left(
        \begin{aligned}
            & + p_2 \cdot (p_3-p_4) (p_5-p_6) \cdot (p_7-p_8)\\
            & -2p_2 \cdot (p_5-p_6) (p_3-p_4) \cdot (p_7-p_8)\\
            & + p_2 \cdot (p_7-p_8) (p_5-p_6) \cdot (p_3-p_4)
        \end{aligned}
        \right)\\\notag
        &-\delta_{12}\times
        \left[
            \frac{\delta_{34}\delta_{56}\delta_{78}}{s_{56}s_{78}s_{5678}s_{345678}} \mathcal{U}(2,5,6,7,8)+\frac{2\delta_{34}\delta_{58}\delta_{67}}{s_{67}s_{5678}s_{345678}} p_2\cdot(p_6-p_7)
            +[45678 \to 84567]
        \right]\\\notag
        &-\delta_{12}\times\left[
        \begin{aligned}
            &\frac{2\delta_{34}\delta_{56}\delta_{78}}{s_{34}s_{56}s_{78}s_{5678}s_{345678}}
            \left(
            \begin{aligned}
            &\mathcal{V}_1(3,4,5,6,7,8)-\mathcal{V}_1(4,3,5,6,7,8)\\
            +&\mathcal{V}_2(3,4,5,6,7,8)-\mathcal{V}_2(3,4,6,5,7,8)\\
            +&\mathcal{V}_2(3,4,7,8,6,5)-\mathcal{V}_2(3,4,8,7,6,5)
            \end{aligned}
            \right)\\
            -&\frac{\delta_{34}\delta_{56}\delta_{78}}{s_{56}s_{78}s_{5678}} \left(\frac{1}{s_{34}}\mathcal{U}(3,5,6,7,8)-\frac{1}{s_{34}}\mathcal{U}(4,5,6,7,8)-\frac{1}{s_{345678}}\mathcal{U}(2,5,6,7,8)\right)\\
            +&
            \left(
              \begin{aligned}
                &-\frac{4\delta_{34}\delta_{58}\delta_{67}}{s_{34}s_{67}s_{5678}s_{345678}} \left(\mathcal{U}(2,3,4,6,7)+\mathcal{W}(3,4,5,8,6,7)\right)\\
                &-\frac{2\delta_{34}\delta_{58}\delta_{67}}{s_{34}s_{67}s_{5678}} (p_3-p_4)\cdot(p_6-p_7)+\frac{2\delta_{34}\delta_{58}\delta_{67}}{s_{67}s_{5678}s_{345678}} p_2\cdot(p_6-p_7)\\
                &+(678 \to 786) \,\,\mathrm{\&}\,\, (567 \to 756)
              \end{aligned}
            \right)\\
            +&[345678 \to 783456]
        \end{aligned}
        \right]\\
        &+\mathrm{mirror},
\end{align}
where $[345678 \to 783456]$ denotes a permutation of the external leg labels, and ``mirror'' represents as before applying $i\to(1-i\mod n)+1$ to all the terms above. Moreover, $(678 \to 786) \,\,\mathrm{\&}\,\, (567 \to 756)$ indicates, for an expression A(5,6,7,8), the operation
\begin{equation}
    A(5,6,7,8)+(678 \to 786) \,\,\mathrm{\&}\,\, (567 \to 756)\equiv A(5,6,7,8)-\frac{1}{2}A(5,7,8,6)-\frac{1}{2}A(7,5,6,8)\,.
\end{equation}
Furthermore, we define
\begin{align}
    \mathcal{U}(2,3,4,7,8)\equiv&
                - p_2 \cdot p_3  (p_3 + 3 p_4) \cdot (p_7 - p_8)
                + p_2 \cdot p_4  (3 p_3 + p_4) \cdot (p_7 - p_8) \notag \\
                &+ p_2 \cdot p_7  (p_7 + 3 p_8) \cdot (p_3 - p_4)
                 - p_2 \cdot p_8  (3 p_7 + p_8) \cdot (p_3 - p_4)\,,
                 \\
    \mathcal{V}_1(3,4,5,6,7,8)\equiv&
    -  p_2 \cdot p_3 (p_3+3p_4) \cdot p_5 (p_5 + 3 p_6) \cdot (p_7 - p_8)
    + p_2 \cdot p_3 (p_3+3p_4) \cdot p_6 (3p_5 + p_6) \cdot (p_7 - p_8) \notag
    \\
    &
    + p_2 \cdot p_3 (p_3+3p_4) \cdot p_7 (p_7 + 3 p_8) \cdot (p_5 - p_6)
    - p_2 \cdot p_3 (p_3+3p_4) \cdot p_8 (3p_7 + p_8) \cdot (p_5 - p_6) \,, 
    \\
    \mathcal{V}_2(3,4,5,6,7,8)\equiv&
     \,\,\, p_2 \cdot p_5 (p_3-p_4) \cdot p_5 (p_5 + 3 p_6) \cdot (p_7 - p_8)
    \notag\\
    &+  p_2 \cdot p_5 (p_3-p_4) \cdot p_6 (5 p_5 \cdot p_7 - 5 p_5 \cdot p_8 + 7 p_6 \cdot p_7 - 7 p_6 \cdot p_8)  \notag
    \\
    &
    +  p_2 \cdot p_5 (p_3-p_4) \cdot p_7 (3 p_5 \cdot p_7 + p_5 \cdot p_8 + 5 p_6 \cdot p_7 - 9 p_6 \cdot p_8) \notag
    \\
    &-  p_2 \cdot p_5 (p_3-p_4) \cdot p_8 ( p_5 \cdot p_7+3 p_5 \cdot p_8 -9 p_6 \cdot p_7 + 5 p_6 \cdot p_8)\,,\\
    \mathcal{W}(3,4,5,6,7,8)\equiv&
    -p_2 \cdot (p_5+p_6) (p_3 - p_4) \cdot (p_7 - p_8)
    +2p_2 \cdot (p_7-p_8) (p_3 - p_4) \cdot (p_5 + p_6)\,.
\end{align}\vspace{1cm}

\noindent \emph{Different Splittings at 8pt}---The following three figures collect diagrams that contribute to the 8-pt amplitude in the $(1\to 3)$, $(1\to 4)$ and $(1\to 5)$ splittings. The diagrams for the odd splittings $(1\to 3)$ and $(1\to 5)$ illustrate explicitly the claim that the splitting route traverses diagonally at least one $V_4$-type vertex, which ensures vanishing of the amplitude on the corresponding kinematic locus diagram by diagram. In case of the $(1\to 3)$ splitting, one can moreover immediately visualize the near-zero factorisation by splitting the diagrams along the red line.

\begin{figure}[h!]
\centering
    \begin{subfigure}[b]{0.2\textwidth}
    \centering
    \includegraphics[width=1\textwidth]{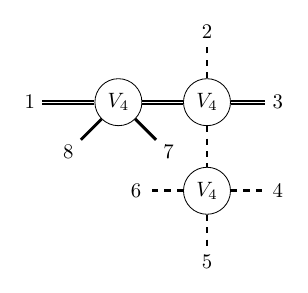}
    \caption{\label{fig:imageac1}}
    \end{subfigure}
    \begin{subfigure}[b]{0.2\textwidth}
    \centering
    \includegraphics[width=1\textwidth]{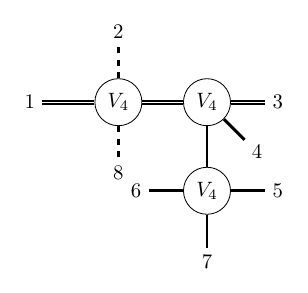}
    \caption{\label{fig:imageac2}}
    \end{subfigure}
    \begin{subfigure}[b]{0.2\textwidth}
    \centering
    \includegraphics[width=1\textwidth]{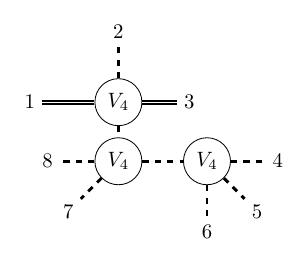}
    \caption{\label{fig:imageac3}}
    \end{subfigure}
    \begin{subfigure}[b]{0.25\textwidth}
    \centering
    \includegraphics[width=1\textwidth]{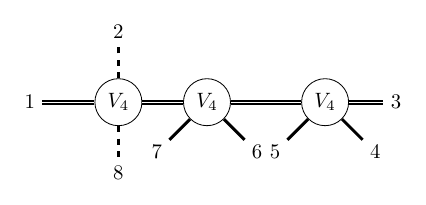}
    \caption{\label{fig:imageac4}}
    \end{subfigure}
    \begin{subfigure}[b]{0.2\textwidth}
    \centering
    \includegraphics[width=1\textwidth]{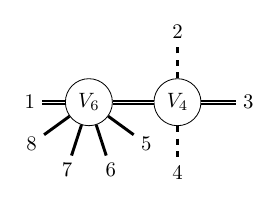}
    \caption{\label{fig:imageac5}}
    \end{subfigure}
    \begin{subfigure}[b]{0.2\textwidth}
    \centering
    \includegraphics[width=1\textwidth]{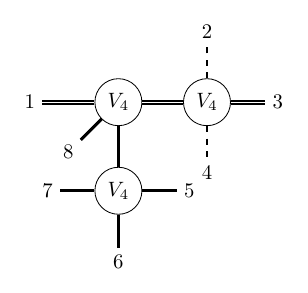}
    \caption{\label{fig:imageac6}}
    \end{subfigure}
    \begin{subfigure}[b]{0.2\textwidth}
    \centering
    \includegraphics[width=1\textwidth]{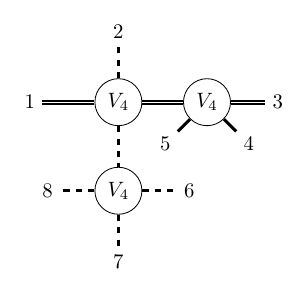}
    \caption{\label{fig:imageac7}}
    \end{subfigure}
    \begin{subfigure}[b]{0.25\textwidth}
    \centering
    \includegraphics[width=1\textwidth]{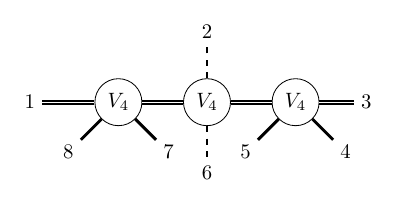}
    \caption{\label{fig:imageac8}}
    \end{subfigure}
    \begin{subfigure}[b]{0.25\textwidth}
    \centering
    \includegraphics[width=1\textwidth]{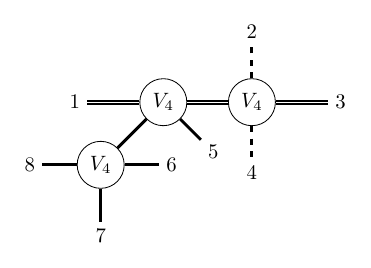}
    \caption{\label{fig:imageac9}}
    \end{subfigure}
    \begin{subfigure}[b]{0.15\textwidth}
    \centering
    \includegraphics[width=1\textwidth]{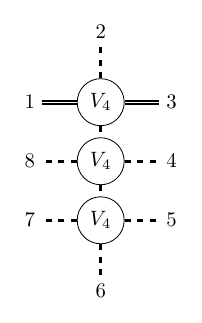}
    \caption{\label{fig:imageac10}}
    \end{subfigure}
    \begin{subfigure}[b]{0.22\textwidth}
    \centering
    \includegraphics[width=1\textwidth]{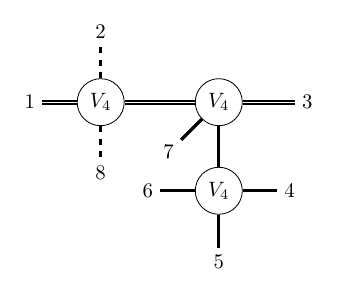}
    \caption{\label{fig:imageac11}}
    \end{subfigure}
    \begin{subfigure}[b]{0.16\textwidth}
    \centering
    \includegraphics[width=1\textwidth]{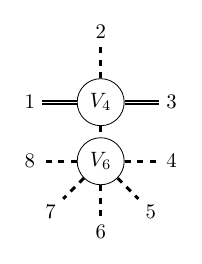}
    \caption{\label{fig:imageac12}}
    \end{subfigure}
    \begin{subfigure}[b]{0.2\textwidth}
    \centering
    \includegraphics[width=1\textwidth]{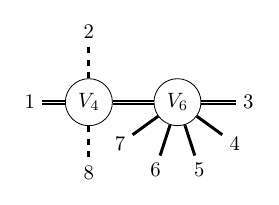}
    \caption{\label{fig:imageac13}}
    \end{subfigure}
    \begin{subfigure}[b]{0.2\textwidth}
    \centering
    \includegraphics[width=1\textwidth]{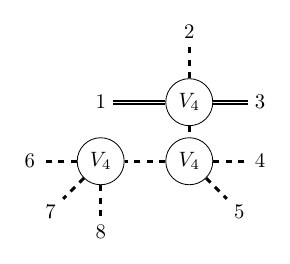}
    \caption{\label{fig:imageac14}}
    \end{subfigure}
    \begin{subfigure}[b]{0.27\textwidth}
    \centering
    \includegraphics[width=1\textwidth]{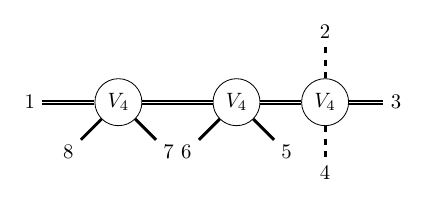}
    \caption{\label{fig:imageac15}}
    \end{subfigure}
\caption{All diagrams of $A_8^{(1\to 3)}$.}
\end{figure}

\begin{figure}[h!]
\centering
    \begin{subfigure}[b]{0.27\textwidth}
    \centering
    \includegraphics[width=1\textwidth]{fig1.pdf}
    \caption{\label{fig:imagead1}}
    \end{subfigure}
    \begin{subfigure}[b]{0.2\textwidth}
    \centering
    \includegraphics[width=1\textwidth]{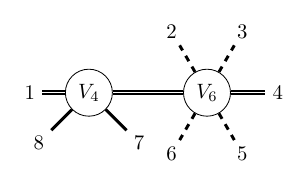}
    \caption{\label{fig:imagead2}}
    \end{subfigure}
    \begin{subfigure}[b]{0.25\textwidth}
    \centering
    \includegraphics[width=1\textwidth]{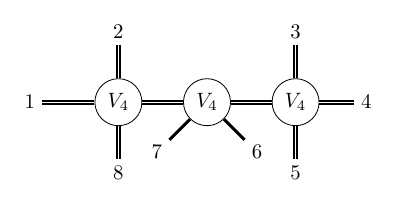}
    \caption{\label{fig:imagead3}}
    \end{subfigure}
    \begin{subfigure}[b]{0.27\textwidth}
    \centering
    \includegraphics[width=1\textwidth]{fig4.pdf}
    \caption{\label{fig:imagead4}}
    \end{subfigure}
    \begin{subfigure}[b]{0.2\textwidth}
    \centering
    \includegraphics[width=1\textwidth]{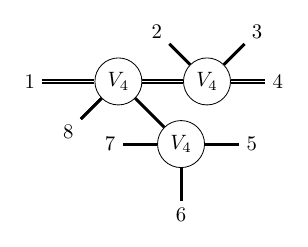}
    \caption{\label{fig:imagead5}}
    \end{subfigure}
    \begin{subfigure}[b]{0.25\textwidth}
    \centering
    \includegraphics[width=1\textwidth]{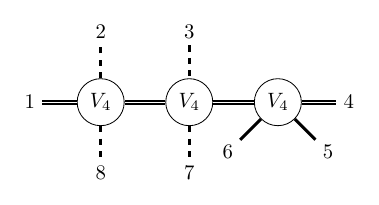}
    \caption{\label{fig:imagead6}}
    \end{subfigure}
    \begin{subfigure}[b]{0.25\textwidth}
    \centering
    \includegraphics[width=1\textwidth]{fig7.pdf}
    \caption{\label{fig:imagead7}}
    \end{subfigure}
    \begin{subfigure}[b]{0.2\textwidth}
    \centering
    \includegraphics[width=1\textwidth]{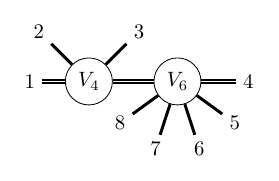}
    \caption{\label{fig:imagead8}}
    \end{subfigure}
    \begin{subfigure}[b]{0.25\textwidth}
    \centering
    \includegraphics[width=1\textwidth]{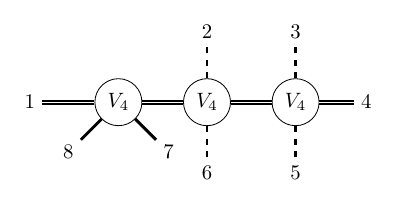}
    \caption{\label{fig:imagead9}}
    \end{subfigure}
    \begin{subfigure}[b]{0.22\textwidth}
    \centering
    \includegraphics[width=1\textwidth]{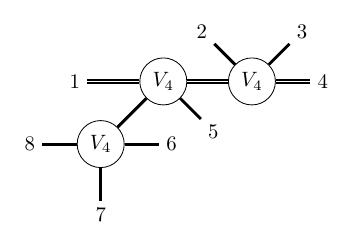}
    \caption{\label{fig:imagead10}}
    \end{subfigure}
    \begin{subfigure}[b]{0.22\textwidth}
    \centering
    \includegraphics[width=1\textwidth]{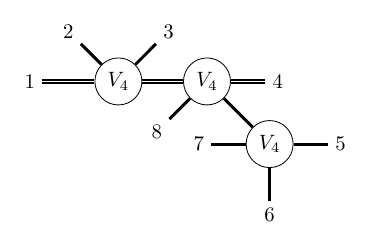}
    \caption{\label{fig:imagead11}}
    \end{subfigure}
    \begin{subfigure}[b]{0.2\textwidth}
    \centering
    \includegraphics[width=1\textwidth]{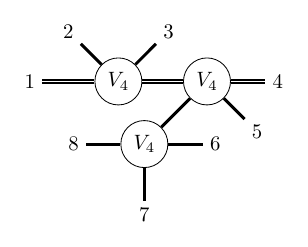}
    \caption{\label{fig:imagead12}}
    \end{subfigure}
    \begin{subfigure}[b]{0.2\textwidth}
    \centering
    \includegraphics[width=1\textwidth]{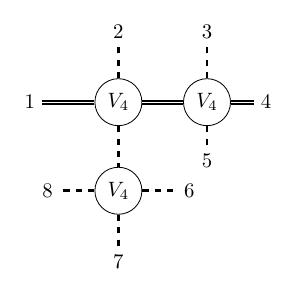}
    \caption{\label{fig:imagead13}}
    \end{subfigure}
    \begin{subfigure}[b]{0.2\textwidth}
    \centering
    \includegraphics[width=1\textwidth]{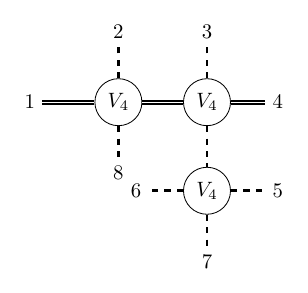}
    \caption{\label{fig:imagead14}}
    \end{subfigure}
    \begin{subfigure}[b]{0.18\textwidth}
    \centering
    \includegraphics[width=1\textwidth]{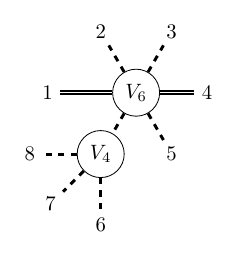}
    \caption{\label{fig:imagead15}}
    \end{subfigure}
    \begin{subfigure}[b]{0.18\textwidth}
    \centering
    \includegraphics[width=1\textwidth]{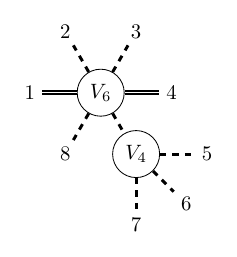}
    \caption{\label{fig:imagead16}}
    \end{subfigure}
    \begin{subfigure}[b]{0.2\textwidth}
    \centering
    \includegraphics[width=1\textwidth]{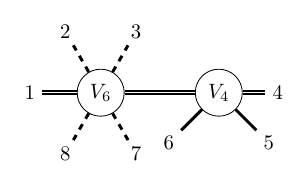}
    \caption{\label{fig:imagead17}}
    \end{subfigure}
    \begin{subfigure}[b]{0.2\textwidth}
    \centering
    \includegraphics[width=1\textwidth]{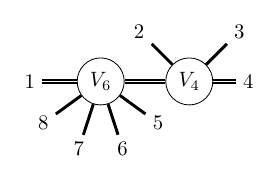}
    \caption{\label{fig:imagead18}}
    \end{subfigure}
    \begin{subfigure}[b]{0.16\textwidth}
    \centering
    \includegraphics[width=1\textwidth]{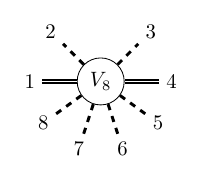}
    \caption{\label{fig:imagead19}}
    \end{subfigure}
\caption{All diagrams of $A_8^{(1\to 4)}$.}
\end{figure}

\begin{figure}[h!]
\centering
    \begin{subfigure}[b]{0.2\textwidth}
    \centering
    \includegraphics[width=1\textwidth]{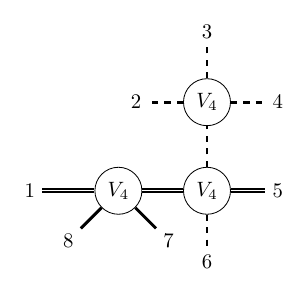}
    \caption{\label{fig:imageae1}}
    \end{subfigure}
    \begin{subfigure}[b]{0.25\textwidth}
    \centering
    \includegraphics[width=1\textwidth]{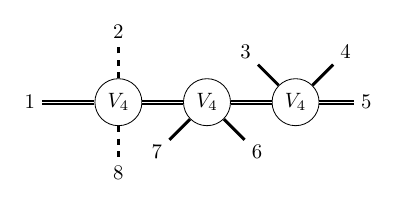}
    \caption{\label{fig:imageae2}}
    \end{subfigure}
    \begin{subfigure}[b]{0.25\textwidth}
    \centering
    \includegraphics[width=1\textwidth]{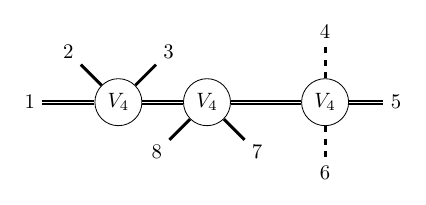}
    \caption{\label{fig:imageae3}}
    \end{subfigure}
    \begin{subfigure}[b]{0.2\textwidth}
    \centering
    \includegraphics[width=1\textwidth]{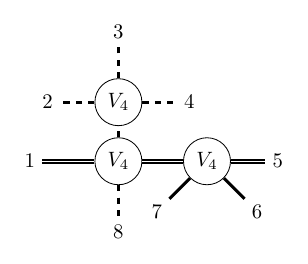}
    \caption{\label{fig:imageae4}}
    \end{subfigure}
    \begin{subfigure}[b]{0.25\textwidth}
    \centering
    \includegraphics[width=1\textwidth]{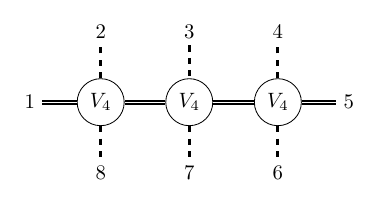}
    \caption{\label{fig:imageae5}}
    \end{subfigure}
    \begin{subfigure}[b]{0.25\textwidth}
    \centering
    \includegraphics[width=1\textwidth]{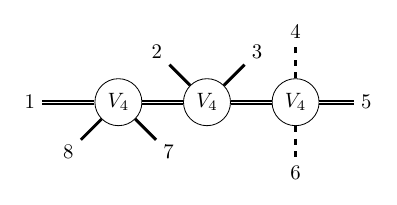}
    \caption{\label{fig:imageae6}}
    \end{subfigure}
    \begin{subfigure}[b]{0.25\textwidth}
    \centering
    \includegraphics[width=1\textwidth]{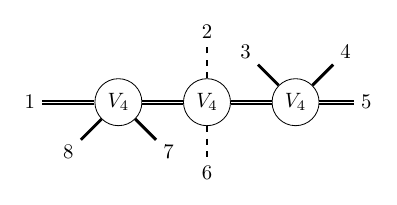}
    \caption{\label{fig:imageae7}}
    \end{subfigure}
    \begin{subfigure}[b]{0.16\textwidth}
    \centering
    \includegraphics[width=1\textwidth]{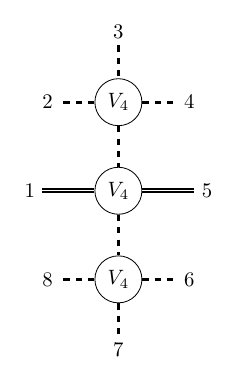}
    \caption{\label{fig:imageae8}}
    \end{subfigure}
    \begin{subfigure}[b]{0.25\textwidth}
    \centering
    \includegraphics[width=1\textwidth]{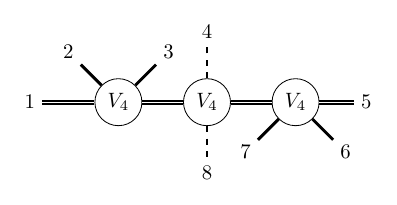}
    \caption{\label{fig:imageae9}}
    \end{subfigure}
    \begin{subfigure}[b]{0.21\textwidth}
    \centering
    \includegraphics[width=1\textwidth]{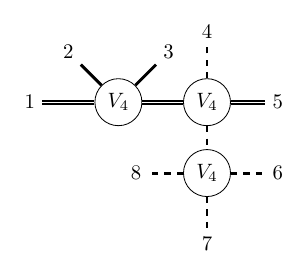}
    \caption{\label{fig:imageae10}}
    \end{subfigure}
    \begin{subfigure}[b]{0.21\textwidth}
    \centering
    \includegraphics[width=1\textwidth]{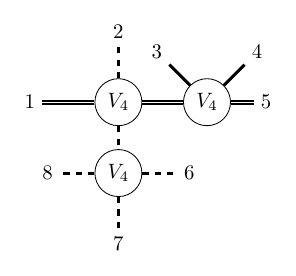}
    \caption{\label{fig:imageae11}}
    \end{subfigure}
    \begin{subfigure}[b]{0.25\textwidth}
    \centering
    \includegraphics[width=1\textwidth]{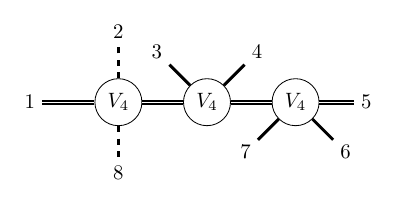}
    \caption{\label{fig:imageae12}}
    \end{subfigure}
    \begin{subfigure}[b]{0.22\textwidth}
    \centering
    \includegraphics[width=1\textwidth]{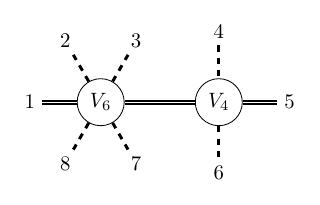}
    \caption{\label{fig:imageae13}}
    \end{subfigure}
    \begin{subfigure}[b]{0.22\textwidth}
    \centering
    \includegraphics[width=1\textwidth]{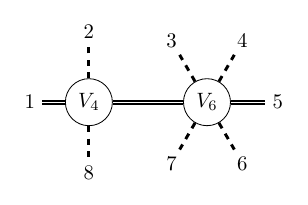}
    \caption{\label{fig:imageae14}}
    \end{subfigure}
\caption{All diagrams of $A_8^{(1\to 5)}$.}
\end{figure}

\end{document}